\documentclass[11pt,a4paper]{article}%
\usepackage{amsfonts}
\usepackage{amsmath}
\usepackage{amssymb}
\usepackage[margin=1in,a4paper]{geometry}
\usepackage[sc,center,compact,noindentafter,medium]{titlesec}
\usepackage[onehalfspacing,nodisplayskipstretch]{setspace}
\usepackage{graphicx}
\usepackage{color}
\usepackage{booktabs}
\usepackage{comment}%
\providecommand{\U}[1]{\protect\rule{.1in}{.1in}}

\definecolor{darkblue}{rgb}{.2, 0.2,.8}
\definecolor{darkgreen}{rgb}{0,0.5,0.3}
\definecolor{darkred}{rgb}{.8, .1,.1}

\providecommand{\U}[1]{\protect\rule{.1in}{.1in}}

\newtheorem{theorem}{\normalfont\scshape Theorem}[section]

\newtheorem{lemma}{\normalfont\scshape Lemma}[section]

\newtheorem{algorithm}{\normalfont\scshape Algorithm}
\newtheorem{remark}{\normalfont\scshape Remark}[section]

\newtheorem{assumption}{\normalfont\scshape Assumption}
\newtheorem{proposition}[theorem]{\normalfont\scshape Proposition}
\expandafter
\def \expandafter \normalsize \expandafter{\normalsize \setlength \abovedisplayskip{10pt plus 2pt minus 7pt}}
\expandafter
\def \expandafter \normalsize \expandafter{\normalsize \setlength \abovedisplayshortskip{0pt plus 2pt}}
\expandafter
\def \expandafter \normalsize \expandafter{\normalsize \setlength \belowdisplayskip{10pt plus 2pt minus 7pt}}
\expandafter
\def \expandafter \normalsize \expandafter{\normalsize \setlength \belowdisplayshortskip{5pt plus 2pt minus 3pt}}

\numberwithin{equation}{section}
\allowdisplaybreaks
\newcounter{archcounter}
\newtheorem{archassumption}[archcounter]{\normalfont\scshape Assumption ARCH}
\newcounter{liniidcounter}
\newtheorem{liniidassumption}[liniidcounter]{\normalfont\scshape Assumption LinIID}

\begin{document}

\title{ }

\begin{center}
{\LARGE \textsc{uniform critical values for likelihood ratio tests in boundary
problems}}%

\renewcommand{\thefootnote}{}
\footnote{
\hspace{-7.2mm}
$^{a}%
$Department of Economics, University of Bologna, Italy and Department of Economics, University of Exeter, UK.
\newline$^{b}$Department of Economics, University of Colorado, USA.
\newline$^{c}$Department of Economics, University of Copenhagen, Denmark.
\newline Correspondence to: Adam McCloskey, email adam.mccloskey@colorado.edu.
\newline
We thank Pascal Lavergne and Frank Kleibergen for comments and suggestions. We also thank participants at the 2022 CIREQ Montreal Econometrics Conference, the 2023 Robust Econometric Methods in Financial Econometrics Workshop in Copenhagen, the 2023 ICEEE Conference in Cagliari, and the 2024 Toulouse Financial Econometrics Conference for comments.
\newline
McCloskey acknowledges support from the National Science Foundation under Grant SES-2341730.
A.~Rahbek and G.~Cavaliere gratefully acknowledge support from the Independent Research Fund Denmark (DFF Grant 7015-00028) and
the Italian Ministry of University and Research (PRIN 2020 Grant 2020B2AKFW).}
\addtocounter{footnote}{-1}
\renewcommand{\thefootnote}{\arabic{footnote}}%
{\normalsize \vspace{0.1cm} }

{\large \textsc{Giuseppe Cavaliere}}$^{a}${\large \textsc{, Adam McCloskey}%
}$^{b}${\large \textsc{, Rasmus S. Pedersen}}$^{c}$ \linebreak%
{\large \textsc{and Anders Rahbek}}$^{c}$

{\normalsize \vspace{0.2cm}\vspace{0.2cm}}

July 25, 2025\bigskip

\textsc{Abstract}\vspace{-0.15cm}
\end{center}

{\small Limit distributions of likelihood ratio statistics are well-known to
be discontinuous in the presence of nuisance parameters at the boundary of the
parameter space, which lead to size distortions when standard critical values
are used for testing. In this paper, we propose a new and simple way of
constructing critical values that yields uniformly correct asymptotic size,
regardless of whether nuisance parameters are at, near or far from the
boundary of the parameter space. Importantly, the proposed critical values are
trivial to compute and at the same time provide powerful tests in most
settings. In comparison to existing size-correction methods, the new approach
exploits the monotonicity of the two components of the limiting distribution
of the likelihood ratio statistic, in conjunction with rectangular confidence
sets for the nuisance parameters, to gain computational tractability.}

{\small Uniform validity is established for likelihood ratio tests based on
the new critical values, and we provide illustrations of their construction in
two key examples:~(i) testing a coefficient of interest in the classical
linear regression model with non-negativity constraints on control
coefficients, and, (ii) testing for the presence of exogenous variables in
autoregressive conditional heteroskedastic models (ARCH)\ with exogenous
regressors. Simulations confirm that the tests have desirable size and power
properties. A brief empirical illustration demonstrates the usefulness of our
proposed test in relation to testing for spill-overs and ARCH(-X)
effects.}\medskip

\noindent\textsc{Keywords}:\ Likelihood ratio tests; Parameters on the
boundary; Uniform inference.

\section*{Introduction}

\label{sec intro}\textsc{As is now well-known}, the asymptotic distribution of
a likelihood ratio (LR) statistic is discontinuous under the null hypothesis
being tested at points for which a vector nuisance parameter lies on the
boundary of its parameter space. This feature complicates the proper formation
of critical values (CVs) that control the asymptotic size of the LR test
uniformly over the parameter space. In this setting, we propose a simple and
computationally-tractable method of CV formation (Algorithm \ref{algorithm})
that controls the asymptotic size of a test using the standard LR statistic.
The correct uniform asymptotic size of the resulting test (Theorem
\ref{thm:asysize control}) ensures that the test also has good size properties
in finite samples whether the nuisance parameter is at, near or far away from
the boundary of its parameter space.

Although a few recent advances in the literature have now produced hypothesis
tests that are uniformly valid over the nuisance parameter space, our focus in
this paper is to introduce a hypothesis test that is simple to compute even
when the dimension of the vector of nuisance parameter is not small, without
sacrificing much power. To attain this goal, we continue to use a standard LR
statistic because it is one of the most widely-used tests in statistics and is
efficient in finite samples under correct specification when nuisance
parameters are far from the boundary. However, the standard $\chi^{2}$-based
CVs for LR statistics are invalid when nuisance parameters are on or near the
boundary. In contrast, the CVs we propose are based upon the asymptotic
distributions of this test statistic that arise under certain parameter
sequences that drift toward the boundary of the parameter space.

In order to establish uniform asymptotic size control, we derive the null
asymptotic distribution of LR statistics under a comprehensive class of
parameter sequences that drift the nuisance parameter vector toward the
boundary at any sample size-dependent rate. Under a particularly important
rate for establishing asymptotic size, the asymptotic distribution of the LR
statistic depends upon a nuisance parameter vector that cannot be estimated
consistently but can still be \textquotedblleft estimated\textquotedblright%
\ by an asymptotically Gaussian random vector centered at its true value. Our
key insight is that confidence bounds for this latter \textquotedblleft
estimator\textquotedblright\ can be combined with monotonicity properties
inherent to the two components of the asymptotic distribution of the LR
statistic to yield CVs that can be computed via straightforward Monte Carlo
simulation. In order to uniformly control asymptotic size, we make use of a
standard Bonferroni correction to account for the randomness involved in the
\textquotedblleft estimation\textquotedblright\ of the nuisance parameter.

The main theoretical result of our paper (Theorem \ref{thm:asysize control})
proves that our algorithm for CV formation (Algorithm \ref{algorithm})
controls axymptotic size uniformly in a general framework encompassing
numerous and varied applications. Examples include tests of a one-dimensional
sub-vector of the mean in the multivariate Gaussian location model with a
restricted mean vector, tests of a regression coefficient in the linear
regression model when some coefficients have a known sign and tests of
parameters in random coefficients models such as the workhorse empirical
industrial organization model of Berry, Levinsohn and Pakes (1995). We verify
that the conditions of our main theoretical result hold in applications of
tests on regression coefficients and specification tests in ARCH-type models,
the latter being a pervasive example of tests suffering from boundary problems
in the recent literature on conditional volatility models, see, e.g., Francq
and Zako\"{\i}an (2009), Cavaliere, Nielsen and Rahbek (2017), Cavaliere,
Nielsen, Pedersen and Rahbek (2022) and Cavaliere, Perera and Rahbek (2024).
We contribute to this literature by providing the first hypothesis test we are
aware of with proven uniform asymptotic validity for specification testing for
the presence of exogenous variables in ARCH-type models.

Several papers in the literature derive asymptotic distributions of estimators
and test statistics at the boundary of the parameter space; see, e.g., Self
and Liang (1987), Shapiro (1989), Geyer (1994), Silvapulle and Silvapulle
(1995) and Andrews (2001). More recently, Cavaliere et al. (2022) propose a
bootstrap-based CV construction that implicitly uses an estimator to switch
between the quantiles of the asymptotic distribution at and far away from the
boundary in large samples. However, using CVs that correspond to distributions
at the boundary and/or far away from the boundary does not ensure uniform
asymptotic size control since null distributions can have larger quantiles in
intermediate ranges near the boundary of the parameter space.

This latter fact has been recognized in the literature, leading to inference
methods that control asymptotic null rejection probabilities uniformly across
all parameter sequences that may drift toward the boundary of the parameter
space. In particular,\ Andrews and Guggenberger\ (2009) propose CVs in a
general testing framework that are asymptotically equivalent to
least-favorable CVs which find the maximal quantile of a test statistic's null
asymptotic distribution over the nuisance parameter that cannot be
consistently estimated. Recent alternative approaches allowing for quite
general constraints, including Hong and Li (2020) and Li (2025), advocate
similar least favorable-type CVs that are derived from sophisticated bootstrap
implementations. Recognizing the conservative nature of tests using these CVs,
and its negative consequence on power, McCloskey (2017) proposes CVs that do
not maximize these quantiles over the entire nuisance parameter space but
rather a first stage confidence set for the nuisance parameter; see also
Berger and Boos (1994) and Silvapulle (1996) for parametric finite-sample
versions of this approach. In the context of boundary problems, Mitchell,
Allman and Rhodes (2019) apply McCloskey's (2017) approach to CV formation for
LR statistics although, unlike the current paper, they focus on problems that
may also feature singularities (see Drton, 2009). Fan and Shi (2023) also take
this approach to CV formation for Wald and Quasi-Likelihood Ratio (QLR)
statistics in more complicated boundary problems that require an initial first
step of identifying an implicit nuisance parameter.

Although McCloskey's (2017) method tends to produce better power properties
than the least favorable approach of Andrews and Guggenberger (2009) and
related papers, both methods can become computationally intractable when the
nuisance parameter is not low-dimensional because they generally require the
optimization of a function whose values must be simulated at each parameter
value. The approach we propose in this paper is similar in spirit to the
approach of McCloskey (2017), but by exploiting the monotonicity properties of
the two components of the LR statistic, remains computationally tractable in
the presence of nuisance parameters that are not low-dimensional because it
does not require optimization of a function calculated via simulation.

Finally, Ketz (2018) and Ketz and McCloskey (2023) propose computationally
tractable and uniformly valid inference approaches to inference when
parameters may be on or near the boundary of the parameter space. Although
these approaches are different from that taken in this paper, one commonality
these two works share with our approach here is that they use one-step
estimators, discussed e.g., in Newey and McFadden (1994), to attain
asymptotically Gaussian estimation as input to form valid inference regardless
of where the true parameter lies in relation to the boundary of their
parameter space.\medskip

\noindent\textsc{Structure of the paper}. The remainder of the paper is
organized as follows. Section \ref{sec:intuition} provides basic intuition for
our proposed method. Section \ref{sec:general_setup} introduces the general
setting and assumptions as well as the ongoing examples on hypothesis testing
in a linear regression model with positivity constrained coefficients and
specification testing in ARCH models. Section \ref{sec:CV} contains our
proposed algorithm for critical value construction and proofs the validity of
the critical values. Section \ref{sec:sim} considers the performance of our
proposed method by means of Monte Carlo simulations, and Section
\ref{sec:emp_illustration} contains a small empirical illustration in relation
to specification testing in ARCH models for various stock market indices. All
proofs and additional simulations are collected in the supplemental appendix.

\bigskip

\noindent\textsc{Notation}. The following notation is used throughout the
paper. The set of positive definite matrices with eigenvalues bounded below by
some $\kappa>0$, and above by $\kappa^{-1}$ is denoted by $\Upsilon$. For a
square $m\times m$ matrix $A$, let $\operatorname*{diag}(A)$ a diagonal matrix
with the same diagonal of $A$. Moreover, let $\operatorname*{diagv}(A)$ denote
the $m$-dimensional column vector with the diagonal elements of $A$ as
entries. $\mathbb{I()}$ is the indicator function. With $A$ a positive
definite matrix and $x$ a vector, $\Vert x\Vert_{A}=\sqrt{x^{\prime}Ax}$.

\section{Main ideas in a simple Gaussian model\label{sec:intuition}}

To obtain the basic intuition underlying our proposed CV construction,
consider the case of a scalar parameter of interest $\gamma$ and a scalar
nuisance parameter $\beta$, where $\gamma\geq0$ and $b\geq0$. We consider
testing $\mathsf{H}_{0}:\gamma=0$, against the alternative, $\mathsf{H}%
_{1}:\gamma>0$. For simplicity, we frame this in terms of the a bivariate
normal model for $Y\in\mathbb{R}^{2}$, where%
\[
Y=\left(  Y_{\gamma},Y_{b}\right)  ^{\prime}\sim\operatorname*{N}\left(
\lambda,\Sigma\right)  ,
\]
with $\lambda=\left(  \gamma,b\right)  ^{\prime}$ and $\Sigma$ a known
positive definite covariance matrix. In standard settings, this model
corresponds to a limiting experiment. For testing $\mathsf{H}_{0}$, consider
the standard likelihood ratio (LR) statistic as defined by%
\[
\mathsf{LR}=\inf_{\lambda\in\{0\}\times\lbrack0,\infty]}(\lambda-Y)^{\prime
}\Sigma^{-1}(\lambda-Y)-\inf_{\lambda\in\lbrack0,\infty]\times\lbrack
0,\infty]}(\lambda-Y)^{\prime}\Sigma^{-1}(\lambda-Y).
\]
Under $\mathsf{H}_{0}$, by standard arguments%
\begin{equation}
\mathsf{LR}\sim\inf_{\lambda\in\{0\}\times\lbrack-b,\infty]}Q(\lambda
)-\inf_{\lambda\in\lbrack0,\infty]\times\lbrack-b,\infty]}Q(\lambda
)\equiv\mathcal{L}(b,b), \label{eq:null dist}%
\end{equation}
for $Q(\lambda)=(\lambda-Z)^{\prime}\Sigma^{-1}(\lambda-Z)$ with
$Z\sim\operatorname*{N}(0,\Sigma)$.

The difficulty with controlling the size of a test using the LR statistic in
this setting comes down to the fact that $\beta$ is unknown and therefore the
null distribution $\mathcal{L}(b,b)$ in (\ref{eq:null dist}) is unknown. Since
the null distribution is unknown, it is not obvious how to define a critical
value (CV) such that the test has size no greater than a nominal level
$\alpha\in\left(  0,1\right)  $. A naive approach which sets the CV to the
$1-\alpha$ quantile of (\ref{eq:null dist}) at some given value of $b$ can
lead to size distortions. For example, the standard CV is equal to the
$1-\alpha$ quantile of (\ref{eq:null dist}) for $b=\infty$, corresponding to a
$\max\{\operatorname*{N}(0,1),0\}^{2}$-distribution. However, a test using
this CV will over-reject if the $1-\alpha$ quantile of the distribution in
(\ref{eq:null dist}) is larger for some value of $b<\infty$. To illustrate
this, the figure below contains the 95-percentile of $\mathcal{L}(b,b)$ for
the case where $\Sigma$ is a correlation matrix with correlation parameter
$\rho$.

\begin{center}%
\hspace*{0mm}\includegraphics[width=0.9\linewidth
]{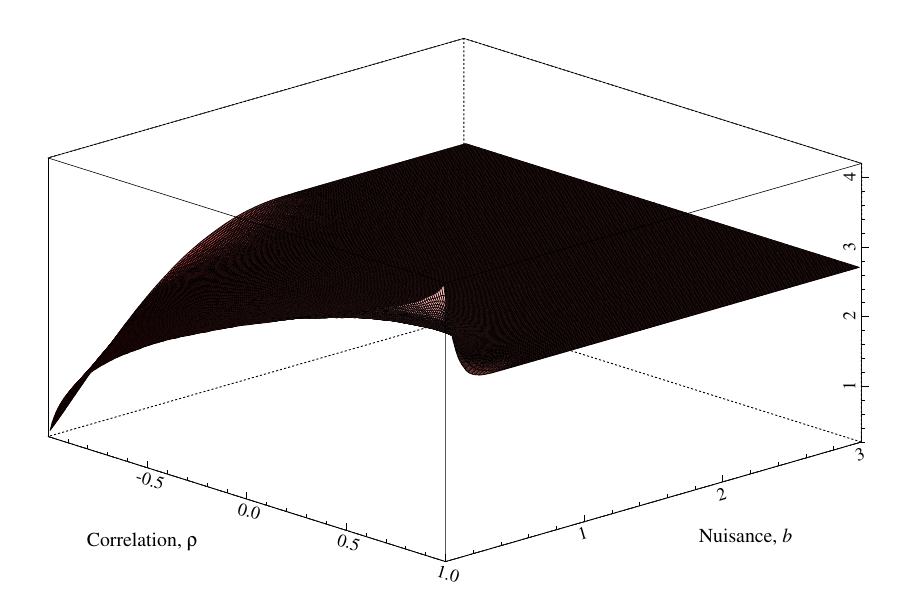}%

{\small Figure: Simulated 95\%-quantiles of }$\mathcal{L}(b,b)$ as a function
of $\rho\in(-1,1)$ and $b\geq0$.
\end{center}

A solution, referred to as the \textquotedblleft least favorable
approach\textquotedblright\ of Andrews and Guggenberger (2009), is based on
using the largest the $1-\alpha$ quantile of (\ref{eq:null dist}) across all
possible values of $b\in\lbrack0,\infty]$ and therefore has correct size.
Although not a major concern in the present example for which $b$ is
one-dimensional, the least favorable approach becomes computationally
intractable as the dimension of the nuisance parameter grows beyond two or so.
Furthermore, this approach can be very conservative, leading to poor power
when the true value of $b$ is not close to the value used to compute the CV.

Our alternative CV is constructed as follows. Define
\[
\mathcal{L(}x,y\mathcal{)=}\inf_{\lambda\in\{0\}\times\lbrack-x,\infty
]}Q(\lambda)-\inf_{\lambda\in\lbrack0,\infty]\times\lbrack-y,\infty]}%
Q(\lambda),
\]
where $\mathcal{L}\left(  x,y\right)  $ is stochastically decreasing in $x$,
while increasing in $y$. Furthermore, let $\mathsf{CV}_{q}(x,y)$ denote the
$q$th quantile of $\mathcal{L(}x,y\mathcal{)}$ and note: (i) under
$\mathsf{H}_{0}$, $\mathsf{LR}\sim\mathcal{L}(b,b)$, and, (ii) for any values
$b_{L}<b_{U}$ with $b\in(b_{L},b_{U})$, $\mathsf{CV}_{q}(b,b)\leq
\mathsf{CV}_{q}(b_{L},b_{U})$. We use these observations to choose $b_{L}$ and
$b_{U}$ in a judicious data-dependent manner to feasibly control the size of
the LR test. Specifically, choose $\eta\in(0,\alpha)$ and let $z_{1-\eta/2}$
denote the $(1-\eta/2)^{th}$ quantile of $\operatorname*{N}(0,1)$. The with
$\Sigma_{22}$ second diagonal entry of $\Sigma$ and
\begin{equation}
b_{L}=Y_{\beta}-\sqrt{\Sigma_{22}}z_{1-\eta/2},\text{ \ \ }b_{U}=Y_{\beta
}+\sqrt{\Sigma_{22}}z_{1-\eta/2}, \label{eq:b bounds}%
\end{equation}
it holds that $\mathbb{P}(b\in(b_{L},b_{u}))=1-\eta$. Thus, our proposal is to
use $\mathsf{CV}_{\alpha}=\mathsf{CV}_{1-\alpha+\eta}(b_{L},b_{U})$ as a CV.
To see how it controls size, note that for any $b\geq0$ the probability of
rejecting under $\mathsf{H}_{0}$ can be bounded above using Bonferroni's
inequality:
\[
\mathbb{P}\left(  \mathcal{L}(b,b)\geq\mathsf{CV}_{\alpha}\right)
\leq\mathbb{P}\left(  \mathsf{CV}_{1-\alpha+\eta}(b,b)\geq\mathsf{CV}_{\alpha
}\right)  +\mathbb{P}\left(  \mathcal{L}(b,b)\geq\mathsf{CV}_{1-\alpha+\eta
}(b,b)\right)  .
\]
The sum of the two terms on the right hand side of the above bound is less
than $\alpha$ since
\[
\mathbb{P}\left(  \mathsf{CV}_{1-\alpha+\eta}(b,b)\geq\mathsf{CV}_{\alpha
}\right)  =1-\mathbb{P}\left(  \mathsf{CV}_{1-\alpha+\eta}(b,b)<\mathsf{CV}%
_{\alpha}\right)  \leq1-\mathbb{P}(b\in(b_{L},b_{U}))=\eta,
\]
and $\mathbb{P}(\mathcal{L}(b,b)\geq\mathsf{CV}_{1-\alpha+\eta}(b,b))=\alpha
-\eta$.

As we demonstrate in Theorem \ref{thm:asysize control}, $\mathsf{CV}_{\alpha}$
can be generalized to control size as a CV for \textsf{$LR$} statistics when
(a) the parameter space for $\gamma$ is not constrained, (b) the dimension of
$\beta$ exceeds one, and, (c) (the estimators) $Y_{\gamma}$ and $Y_{b}$ are
not normally distributed in finite samples. For (a), the minimization space
$[0,\infty]$ in the expressions for \textsf{$LR$} and $\mathcal{L}(b,b)$ are
modified to properly reflect the parameter space for $\gamma$ under
$\mathsf{H}_{1}$. For (b), in order to use the multivariate version of the
monotonicity property (ii), $b_{L}$ and $b_{U}$ are replaced by multivariate
counterparts derived from rectangular confidence sets for the mean of a
multivariate normal distribution. For (iii), the CVs described in Section
\ref{sec:CV} below do not require normally distributed estimators but rather
estimators that are asymptotically Gaussian. We describe how to obtain such
estimators in the presence of boundary constraints via a one-step
Newton-Raphson iteration below.

\section{General setup and framework\label{sec:general_setup}}

\subsection{Setup}

For a given set of observations $\{W_{t}\}_{t=1}^{n}$, consider the
(log-likelihood) objective function,%
\begin{equation}
L_{n}(\theta)=f_{n}(\{W_{t}\}_{t=1}^{n};\theta), \label{Likelihood}%
\end{equation}
in terms of the parameter vector $\theta=(\gamma^{\prime},\beta^{\prime
},\delta^{\prime})^{\prime}\in\Theta=\Theta_{\gamma}\times\Theta_{\beta}%
\times\Theta_{\delta}$, where $\gamma\in\mathbb{R}^{d_{\gamma}}$ is the
parameter (vector) of interest, while $\beta\in\mathbb{R}^{d_{\beta}}$ and
$\delta\in\mathbb{R}^{d_{\delta}}$ are nuisance parameters. Specifically, we
throughout assume that the true value of $\delta$ be an interior point of
$\Theta_{\delta}$, while the true value of $\beta$ is allowed to be in the
interior, or at (near) the boundary of $\Theta_{\beta}$.\textbf{ }Furthermore,
for $\gamma$ we assume $s$ of the components are in the "interior", while the
remaining $d_{\gamma}-s$ are allowed to be on the "boundary", $0\leq s\leq
d_{\gamma}$. Formally (without loss of generality), $\Theta_{\gamma}%
=[\gamma_{L},\gamma_{U}]^{s}\times\lbrack0,\gamma_{U}]^{d_{\gamma}-s}$,
$\Theta_{\beta}=[0,\beta_{U}]^{d_{\beta}}$, with $-\infty\leq\gamma
_{L}<0<\gamma_{U}\leq\infty$, $0<\beta_{U}\leq\infty$, and $\Theta_{\delta
}\subset(-\infty,\infty)^{d_{\delta}}$ is a compact subset. Finally, as in
Andrews and Cheng (2012), we introduce an additional parameter $\phi$ used
here to capture any features of the distribution of the data $\left\{
W_{t}\right\}  $ not explicit from the formulation in (\ref{Likelihood}). This
way, $\psi=(\theta,\phi)$ `completely determines the distribution of the
data'. The parameter space is $\Psi=\{\psi=(\theta,\phi):\theta\in\Theta
,\phi\in\Phi(\theta)\}$ where $\Phi(\theta)\subset\Phi$ with $\Phi$ a
compact\ metric space which induces weak convergence of the bivariate
distributions $(W_{t},W_{t+s})$, all $t,s\geq1$.

We are interested in testing the hypothesis%
\[
\mathsf{H}_{0}:\gamma=\gamma_{0}%
\]
by using the (quasi-) LR statistic. That is, with the unrestricted
$\hat{\theta}_{n}$ and restricted $\tilde{\theta}_{n}$ estimators as defined
by
\[
\hat{\theta}_{n}=\arg\max_{\theta\in\Theta}L_{n}(\theta)\text{ \ \ and
\ \ }\tilde{\theta}_{n}=\arg\max_{\theta\in\Theta_{\mathsf{H}_{0}}}%
L_{n}(\theta),
\]
where $\Theta_{\mathsf{H}_{0}}=\{\theta=(\gamma^{\prime},\beta^{\prime}%
,\delta^{\prime})^{\prime}\in\Theta:\gamma=\gamma_{0}\}$, the statistic is
given by,%
\begin{equation}
\mathsf{LR}_{n}=2(L_{n}(\hat{\theta}_{n})-L_{n}(\tilde{\theta}_{n})).
\label{eq:def:LR}%
\end{equation}

To provide context and intuition, we include as running examples linear
regression models with sign-restricted coefficients as well as a time series
ARCH model with explanatory covariates.

\bigskip

\noindent\textsc{Running example: Regression. }With $W_{t}=(y_{t}%
,x_{t}^{\prime})^{\prime}$ consider the regression equation,%
\begin{equation}
y_{t}=\theta^{\prime}x_{t}+\varepsilon_{t},\text{ for }t=1,2,\ldots,n,
\label{eq:lin_regression}%
\end{equation}
with $x_{t}=(x_{1,t},x_{2,t}^{\prime})^{\prime}$, $\theta=(\gamma
,\beta^{\prime})^{\prime}\in\Theta=\Theta_{\gamma}\times\Theta_{\beta
}=[0,\infty]^{1+d_{\beta}}$. We are interested in testing $\mathsf{H}%
_{0}:\gamma=0$, which we emphasize is a non-standard testing problem as some
of entries of $\beta$ may be zero-valued such that $\beta$ is a boundary point
of $\Theta_{\beta}$.

Given a sample $\left\{  W_{t}\right\}  _{t=1}^{n}$, the least-squares
objective function is given by%
\[
L_{n}(\theta)=-\tfrac{1}{2}\sum_{t=1}^{n}\left(  y_{t}-x_{t}^{\prime}%
\theta\right)  ^{2}=-\tfrac{n}{2}\left(  S_{yy}+\theta^{\prime}S_{xx}%
\theta-2\theta^{\prime}S_{xy}\right)  ,
\]
with $S_{yy}=n^{-1}\sum_{t=1}^{n}y_{t}^{2}$, $S_{xx}=n^{-1}\sum_{t=1}^{n}%
x_{t}x_{t}^{\prime}$, and $S_{xy}=n^{-1}\sum_{t=1}^{n}x_{t}y_{t}$. With the
ordinary least-squares estimator given by $\hat{\theta}_{LS}=S_{xx}^{-1}%
S_{xy}$, it follows that the unrestricted estimator is given by
\[
\hat{\theta}_{n}=\arg\min_{\theta\in\Theta}(\theta-\hat{\theta}_{LS})^{\prime
}S_{xx}(\theta-\hat{\theta}_{LS})\text{,}%
\]
see Lemma \ref{lem:nnls} in the Appendix, while the restricted estimator is
given by%
\[
\tilde{\theta}_{n}=\arg\min_{\theta\in\Theta_{\mathsf{H}_{0}}}(\theta
-\hat{\theta}_{LS})^{\prime}S_{xx}(\theta-\hat{\theta}_{LS}),
\]
where $\Theta_{\mathsf{H}_{0}}=\{\theta=(\gamma,\beta^{\prime})^{\prime}%
\in\Theta:\gamma=0\}$, and, finally, the LR statistic is given by
(\ref{eq:def:LR}).

Finally, we note that, in this example, the additional parameter $\phi$
determines the joint distribution of $x_{t}$ and $\varepsilon_{t}$%
.\hfill$\square$

\bigskip

\noindent\textsc{Running example: ARCH. }For the second example, we consider
hypothesis testing in ARCH models augmented with non-negative explanatory (or
exogenous, X) covariates. Let $y_{t}\in%
\mathbb{R}
$ be given by
\begin{align}
y_{t}  &  =\sigma_{t}\varepsilon_{t},\quad t\in%
\mathbb{Z}
,\label{eq:ARCHX_y}\\
\sigma_{t}^{2}  &  =\theta^{\prime}F_{t-1} \label{eq:sigma}%
\end{align}
where $F_{t}=\left(  X_{t}^{\prime},Y_{t}^{\prime},1\right)  ^{\prime}$, with
$X_{t}=(x_{1,t},\dots,x_{p,t})^{\prime}$ and $Y_{t}=(y_{t}^{2},\dots
y_{t-q+1}^{2})^{\prime}$. We are interested in testing whether the covariates
are needed for the conditional variance $\sigma_{t}^{2}$. With $\theta$
partitioned as $\theta=\left(  \gamma^{\prime},\beta^{\prime},\delta\right)
^{\prime}$, let $F_{t}=(Y_{\gamma,t}^{\prime},Y_{\beta,t}^{\prime},1)^{\prime
}$, write (\ref{eq:sigma}) as a function of $\theta$ as%
\[
\sigma_{t}^{2}\left(  \theta\right)  =\theta^{\prime}F_{t-1}=\delta
+\beta^{\prime}Y_{\beta,t-1}+\gamma^{\prime}Y_{\gamma,t-1}\text{,}%
\]
with $\theta\in\Theta=\Theta_{\gamma}\times\Theta_{\beta}\times\Theta_{\delta
}$, $\Theta_{\gamma}=\left[  0,\gamma_{U}\right]  ^{p}$, $\Theta_{\beta
}=\left[  0,\gamma_{U}\right]  ^{q}$ and $\Theta_{\delta}=[\delta_{L}%
,\delta_{U}]$, $\delta_{L}>0$. As before, the objective is to test the
hypothesis $\mathsf{H}_{0}:\gamma=0$, which is a non-standard testing problem
as some of the $\beta$'s can be on the boundary.

Given a sample $\left\{  W_{t}\right\}  _{t=1}^{n}=\{(y_{t}^{2},F_{t-1}%
^{\prime})\}_{t=1}^{n}$, the Gaussian quasi-log-likelihood function is (up to
a constant) given by%
\begin{equation}
L_{n}(\theta)=\sum_{t=1}^{n}l_{t}\left(  \theta\right)  ,\text{ \ \ \ }%
l_{t}(\theta)=-\frac{1}{2}\left(  \log\sigma_{t}^{2}(\theta)+\frac{y_{t}^{2}%
}{\sigma_{t}^{2}(\theta)}\right)  \text{.} \label{eq:ARCH-likelihood}%
\end{equation}
The QMLE $\hat{\theta}_{n}$ ($\tilde{\theta}_{n}$) is any maximizer of
$L_{n}\left(  \theta\right)  $ over $\Theta$ ($\Theta_{\mathsf{H}_{0}%
}=\{\theta\in\Theta:\gamma=0\}$), and%
\begin{equation}
\mathsf{LR}_{n}(\mathsf{H}_{0})=2\left[  L_{n}(\hat{\theta}_{n})-L_{n}%
(\tilde{\theta}_{n})\right]  \label{eq:ARCH_LR_stat}%
\end{equation}
is the LR\ statistic.

Finally, in this example the additional parameter $\phi=\phi(\theta)$
determines the joint distribution of $F_{t-1}$ and $\varepsilon_{t}$%
.\hfill$\square$

\subsection{Drifting sequences and assumptions}

In order to establish uniform validity of our proposed critical values in the
possible presence of nuisance parameters near or on the boundary, we study the
limiting behavior of the LR statistic in (\ref{eq:def:LR}) under drifting
sequences of parameters $\psi_{n}=(\theta_{n},\phi_{n})\rightarrow\psi
_{0}=(\theta_{0},\phi_{0})$. Specifically, we let $\theta_{n}=\left(
\gamma_{0}^{\prime},\beta_{n}^{\prime},\delta_{n}^{\prime}\right)  ^{\prime}$
denote a sequence of parameters in $\Theta_{\mathsf{H}_{0}}$, that is, a
sequence along which $\mathsf{H}_{0}$ holds and which satisfy $\theta
_{n}\rightarrow\theta_{0}=\left(  \gamma_{0}^{\prime},\beta_{0}^{\prime
},\delta_{0}^{\prime}\right)  ^{\prime}$ as $n\rightarrow\infty$. In order to
allow $\beta$ to be either an interior point or near/at the boundary, in
addition to $\beta_{n}=\left(  \beta_{n,1},\ldots,\beta_{n,d_{\beta}}\right)
^{\prime}\rightarrow\beta_{0}$ we require that $\sqrt{n}\beta_{n}\rightarrow
b=(b_{1},b_{2},\dots,b_{d_{\beta}})^{\prime}\in\lbrack0,\infty]^{d_{\beta}}$,
as $n\rightarrow\infty$. This includes sequences of true parameters converging
to an interior point -- e.g., $b_{i}=\infty$ -- or to a boundary point at the
standard $\sqrt{n}$ rate -- e.g., $b_{i}\in\lbrack0,\infty)$. Moreover,
parameters converging to zero at a rate slower (faster) than $\sqrt{n}$
corresponds to, e.g., $b_{i}=\infty$ ($b_{i}=0$). Henceforth, the distribution
of the (stationary) random vectors $\{W_{t}:t\geq1\}$ is determined by the
true parameter $\psi$; expectations, variances and probabilities computed
under $\mathcal{\psi}$ are denoted as $\mathbb{E}_{\psi}$, $\mathbb{V}_{\psi}$
and $\mathbb{P}_{\psi}$, respectively.

In order to state the limiting distribution of $\mathsf{LR}_{n}$, we make the
following assumptions which are similar to Andrews (2001); see also Ketz
(2018) and Fan and Shi (2023). These assumptions are stated for any drifting
sequence $\psi_{n}\in\Psi$ satisfying%
\begin{equation}
\psi_{n}\rightarrow\psi_{0}\quad\text{and\quad}\sqrt{n}\beta_{n}\rightarrow
b=(b_{1},b_{2},\dots,b_{d_{\beta}})^{\prime}\in\lbrack0,\infty]^{d_{\beta}%
}\text{.} \label{eq:def_drifting_sequence}%
\end{equation}

\begin{assumption}
\label{ass:consistency} $\hat{\theta}_{n}-\theta_{n}=o_{p}(1)$ and
$\tilde{\theta}_{n}-\theta_{n}=o_{p}(1)$ as $n\rightarrow\infty$.
\end{assumption}

\begin{assumption}
\label{ass:derivatives}\mbox{} (i)$\ L_{n}(\theta)$ has continuous left/right
partial derivatives of order two on $\Theta$ for all $n\geq1$ (almost surely);
(ii) for all deterministic $\epsilon_{n}\rightarrow0$,
\[
\sup_{\theta\in\Theta:\Vert\theta-\theta_{n}\Vert\leq\epsilon_{n}}\left\Vert
n^{-1}\frac{\partial^{2}L_{n}(\theta)}{\partial\theta\partial\theta^{\prime}%
}-n^{-1}\frac{\partial^{2}L_{n}(\theta_{n})}{\partial\theta\partial
\theta^{\prime}}\right\Vert =o_{p}(1).
\]

\end{assumption}

\begin{assumption}
\label{ass:hessian} The Hessian satisfies%
\[
n^{-1}\frac{\partial^{2}L_{n}(\theta_{n})}{\partial\theta\partial
\theta^{\prime}}\overset{p}{\rightarrow}-\Omega_{0}\in \Upsilon.
\]

\end{assumption}

\begin{assumption}
\label{ass:score}The score satisfies%
\[
n^{-1/2}\frac{\partial L_{n}(\theta_{n})}{\partial\theta}%
\overset{d}{\rightarrow}N(0,\Sigma_{0}),\quad\text{with }\Sigma_{0}\in \Upsilon
\]

\end{assumption}

\begin{remark}
\label{rem:convergence_rate}Note that Assumptions \ref{ass:consistency}%
--\ref{ass:score} imply that $(\hat{\theta}_{n}-\theta_{n})$ and
$(\tilde{\theta}_{n}-\theta_{n})$ are $O_{p}(n^{-1/2})$ (by an extension of
Andrews, 1999, Lemma 1 and Theorem 1).
\end{remark}

\bigskip

\noindent\textsc{Running example: Regression. }We consider here a setting
where the data $\left\{  W_{t}\right\}  $ are assumed to be i.i.d.; see Remark
\ref{rem: time series regr} below for the extension to the more general case
of time series data. Specifically, we make throughout the following standard assumptions.

\begin{liniidassumption}
\label{ass:lin-iid}For all $\psi\in\Psi$,

\begin{enumerate}
\item $\{(x_{t}^{\prime},\varepsilon_{t})^{\prime}\}_{t=1,2,\dots}$ are
i.i.d., with $\mathbb{E}_{\psi}[\varepsilon_{t}|x_{t}]=0$ almost surely for
all $t$;

\item $\mathbb{E}_{\psi}[x_{t}x_{t}^{\prime}]\in \Upsilon$ and $\mathbb{E}%
_{\psi}[x_{t}x_{t}^{\prime}\varepsilon_{t}^{2}]\in \Upsilon$,

\item $\mathbb{E}_{\psi}|x_{j,t}|^{2+\nu}\leq c$ and $\mathbb{E}_{\psi
}|x_{j,t}\varepsilon_{t}|^{2+\nu}\leq c$ for $j=1,\dots,d_{\beta}$ and
constants $c<\infty$ and $\nu>0$ (not depending on $\psi$).
\end{enumerate}
\end{liniidassumption}

\noindent Note that under Assumption \ref{ass:lin-iid} and any drifting
sequence $\psi_{n}\in\Psi$ satisfying (\ref{eq:def_drifting_sequence}), it
holds that $\mathbb{E}_{\psi_{n}}[x_{t}x_{t}^{\prime}]\rightarrow
\mathbb{E}_{\psi_{n}}[x_{t}x_{t}^{\prime}]=\Omega_{0}$ and $\mathbb{E}%
_{\psi_{n}}[x_{t}x_{t}^{\prime}\varepsilon_{t}^{2}]\rightarrow\mathbb{E}%
_{\psi_{0}}[x_{t}x_{t}^{\prime}\varepsilon_{t}^{2}]=\Sigma_{0}$. We have the
following result that ensures that the high-level Assumptions
\ref{ass:consistency}--\ref{ass:score} hold for linear regressions.

\begin{proposition}
\label{prop:reg ass verification} Assumption LinIID \ref{ass:lin-iid} implies
that Assumptions \ref{ass:consistency}--\ref{ass:score} hold with $\Omega
_{0}=\mathbb{E}_{\psi_{0}}[x_{t}x_{t}^{\prime}]$ and $\Sigma_{0}%
=\mathbb{E}_{\psi_{0}}[x_{t}x_{t}^{\prime}\varepsilon_{t}^{2}]$.
\hfill$\square$
\end{proposition}

\bigskip

\noindent\textsc{Running example: ARCH. }Recall that $\psi=(\theta,\phi)$
where $\phi$ denotes the joint distribution of $F_{t-1}$ and $\varepsilon_{t}%
$. In particular, under the assumptions below, $\Omega_{0}$ and $\Sigma_{0}$
in Assumptions \ref{ass:hessian}--\ref{ass:score} are determined entirely by
$\psi$.

\begin{archassumption}
\label{ass:ARCH-HL-stat-mixing}For any $\psi\in\Psi$ the process $\{(y_{t}%
^{2},F_{t-1}^{\prime})^{\prime}\}_{t\in%
\mathbb{Z}
}$ is stationary and $\alpha$-mixing.
\end{archassumption}

The assumption is similar to that used in most of the literature on GARCH
models, such as Francq and Thieu (2018) where the DGP\ is assumed to be
strictly stationary and ergodic. We here impose the stronger assumption of
strong mixing, to apply a weak law of large numbers (LLN) for triangular arrays.

To ensure identification, we make the following assumption.

\begin{archassumption}
\label{ass:identification} For any non-zero constant vector $k\in R^{d_{\beta
}+d_{\gamma}}$ and constant $\tilde{k}\in%
\mathbb{R}
$, it holds that%
\[
\mathbb{P}_{\psi}(k^{\prime}(Y_{t}^{\prime},X_{t}^{\prime})^{\prime}\neq
\tilde{k})>0\quad\text{for all }\psi\in\Psi.
\]

\end{archassumption}

\begin{archassumption}
\label{ass:ARCH-HL-innovation}For all $\psi\in\Psi$ the innovation
$\varepsilon_{t}$ is independent of the $\sigma$-field
\begin{equation}
\mathcal{F}_{t-1}=\sigma\{F_{s}:s\leq t-1\}. \label{eq:sigma_field}%
\end{equation}
\noindent Moreover, $\mathbb{E}_{\psi}[\varepsilon_{t}^{2}-1]=0$, and there
exist constants $\nu_{1},c_{1}\in(0,\infty)$ such that $\mathbb{E}_{\psi
}[|\varepsilon_{t}|^{4(1+\nu_{1})}]\leq c_{1}$ for all $\psi\in\Psi$. There
exists a constant $\kappa\in(0,\infty)$ such that $\mathbb{E}_{\psi
}[\varepsilon_{t}^{4}-1]=\kappa<\infty$ for all $\psi\in\Psi$.
\end{archassumption}

\begin{archassumption}
\label{ass:ARCH-HL-criterion} There exist constants $\nu_{2},c_{2}\in
(0,\infty)$ such that%
\[
\mathbb{E}_{\psi}[\left(  y_{t}^{2}\Vert F_{t-1}\Vert^{3}\right)  ^{1+\nu_{2}%
}]\leq c_{2}\quad\text{for all }\psi\in\Psi.
\]

\end{archassumption}

\noindent We note that the space $\Theta$ is constrained such that
$\{(y_{t},F_{t-1}^{\prime})^{\prime}\}_{t\in%
\mathbb{Z}
}$ is stationary for any $\theta_{0}\in\Theta$. \noindent We have the
following result.

\begin{proposition}
\label{prop:ARCH-LR-stat}Assumptions ARCH \ref{ass:ARCH-HL-stat-mixing}%
-\ref{ass:ARCH-HL-criterion} imply that Assumptions \ref{ass:consistency}%
-\ref{ass:score} hold. \hfill$\square$
\end{proposition}

\subsection{The asymptotic distribution of the LR\ statistic}

The following lemma states the limiting distribution of the LR statistic when
some of the nuisance parameters can be on or near the boundary. It extends
existing results in the literature (e.g., Theorem 4 of Andrews, 2001) as we
consider drifting sequences and hence nuisance parameters local to the
boundary of the parameter space. To state the results, define the generic
quadratic form%
\begin{equation}
Q\left(  \lambda\right)  =\Vert\lambda-HZ\Vert_{(H\Omega_{0}^{-1}H^{\prime
})^{-1}}^{2}\text{,} \label{def Quadratic}%
\end{equation}
where $\lambda\in\mathbb{R}^{d_{\gamma}+d_{\beta}}$ and $H$ is a $\left(
d_{\gamma}+d_{\beta}\right)  \times d_{\theta}$ selection matrix such that
$H\theta=(\gamma^{\prime},\beta^{\prime})^{\prime}$. Moreover, $Z$ is
$N(0,\Omega_{0}^{-1}\Sigma_{0}\Omega_{0}^{-1})$ distributed, with the matrices
$\Omega_{0}$ and $\Sigma_{0}$ provided by, respectively, by Assumptions
\ref{ass:hessian} and \ref{ass:score}.

\begin{lemma}
\label{lem: LR}Under Assumptions \ref{ass:consistency}--\ref{ass:score}, and
any sequence $\psi_{n}\in$ $\Psi$ satisfying (\ref{eq:def_drifting_sequence}),%
\[
\mathsf{LR}_{n}\overset{d}{\rightarrow}\mathcal{L}_{\infty}(b,b)
\]
with%
\begin{equation}
\mathcal{L}_{\infty}(x,y)=\inf_{\lambda\in\{0\}^{d_{\gamma}}\times
\Lambda_{\beta}(x)}Q\left(  \lambda\right)  -\inf_{\lambda\in\Lambda_{\gamma
}\times\Lambda_{\beta}(y)}Q\left(  \lambda\right)  , \label{eq:def:L_infty}%
\end{equation}
for $x,y\in\lbrack0,\infty]^{d_{\beta}}$, and%
\[
\Lambda_{\gamma}=\lim_{n\rightarrow\infty}\sqrt{n}(\Theta_{\gamma}-\gamma
_{0}),\text{ \ \ }\Lambda_{\beta}(b)=[-b_{1},\infty]\times\cdots\times
\lbrack-b_{d_{\beta}},\infty]
\]
where $\Lambda_{\gamma}$ is of the form $\Lambda_{\gamma}=\Lambda_{\gamma
,1}\times\cdots\times\Lambda_{\gamma,d_{\gamma}}$ with $\Lambda_{\gamma
,i}=[-\infty,\infty]$ if $\gamma_{0,i}$ is in the interior of $[\gamma
_{L},\gamma_{U}]$ and $\Lambda_{\gamma,i}=[0,\infty]$ if $\gamma_{0,i}%
=\gamma_{L}$, $i=1,\dots,d_{\gamma}$.
\end{lemma}

The distribution of the limiting random variable $\mathcal{L}_{\infty}(b,b)$
approximates the finite-sample distribution of $\mathsf{LR}_{n}$ for
$\beta_{n}=b/\sqrt{n}$. Notice that this distribution is unknown in practice,
as it is unknown whether the nuisance parameters in $\beta$ are interior
points or on/near the boundary. Put differently, it is not feasible to
consistently estimate the quantiles of $\mathcal{L}_{\infty}(b,b)$ for use as
CVs in this context because $b$ is not consistently estimable along
$\{\psi_{n}\}$ sequences due to the $\sqrt{n}$ scaling of $\beta_{n}$.

Indeed, controlling the asymptotic size of a likelihood ratio test using
$\mathsf{LR}_{n}$ as the test statistic requires the use of a CV that
asymptotically bounds the rejection probability under all parameter sequences
$\psi_{n}\rightarrow\psi_{0}$ satisfying (\ref{eq:def_drifting_sequence}) as
$n\rightarrow\infty$ by the nominal level of the test $\alpha\in(0,1)$ (see,
e.g.,~Andrews and Guggenberger, 2009, or McCloskey, 2017). This motivates us
to examine an alternative method of CV construction that can feasibly control
the asymptotic size of the test. This we do next.

\section{Feasible uniform critical value construction\label{sec:CV}}

In this section we detail how to construct uniformly valid critical values for
the LR statistic. As we argue below, these CVs are simple to construct, allow
to control asymptotic size irrespectively of the nuisance parameters being on
the boundary or not, and provide power gains with respect to existing methods.

As outlined in Section \ref{sec:intuition}, and detailed in the next, that as
$\mathcal{L}_{\infty}(b,b)$ is given by (\ref{eq:def:L_infty}) for $x=y=b,$ we
suggest to exploit the properties of $\mathcal{L}_{\infty}(x,y)$ in order to
select $x$ and $y$ in a data-driven (and simple) way such that the asymptotic
size is bounded above by the user-chosen nominal level $\alpha$.

We proceed as follows. In Section \ref{sec monotone} we present a key
monotonicity property of $\mathcal{L}_{\infty}(x,y)$ and discuss initially a
naive (inefficient) method to construct valid CVs. In Section
\ref{sec ingredients} we introduce some preliminaries needed for constructing
our proposed CV. Our main algorithm to construct the CVs is given in Section
\ref{sec our proposal}, where we also prove the uniform validity of our procedure.

\subsection{Monotonicity of $\mathcal{L}_{\infty}(x,y)$ and some naive CVs}

\label{sec monotone}

A key propery of $\mathcal{L}_{\infty}(x,y)$, which we exploit throughout, is
given in the following lemma.

\begin{lemma}
\label{lem:monotonicity}Let $\underline{b},b,\bar{b}\in\lbrack0,\infty
]^{d_{\beta}}$ satisfy (element-wise) $\underline{b}\leq b\leq\bar{b},$ then%
\[
\mathcal{L}_{\infty}(b,b)\leq\mathcal{L}_{\infty}(\underline{b},\bar{b})\text{
}a.s.
\]

\end{lemma}

Given the monotonicity property of Lemma \ref{lem:monotonicity}, a naive and
straightforward way of constructing a uniformly valid test is to use a CV
based on the distribution of $\mathcal{L}_{\infty}(0^{d_{\beta}}%
,\infty^{d_{\beta}})$, which satisfies $\mathcal{L}_{\infty}(b,b)\leq
\mathcal{L}_{\infty}(0^{d_{\beta}},\infty^{d_{\beta}})$. However, such a
choice of CVs leads to a very conservative test. Alternatively, one may use
the $1-\alpha$ quantile of $\mathcal{L}_{\infty}(b^{\ast},b^{\ast})$, where
$b^{\ast}$ is the maximizer across $[0,\infty]^{d_{\beta}}$ of all $1-\alpha$
quantiles of $\mathcal{L}_{\infty}(b,b)$, in Andrews and Guggenberger (2009).
However, also this choice of CV can lead to a conservative test with lower
power. Perhaps more importantly, $b^{\ast}$ is computationally prohibitive to
compute when the number of nuisance parameters $d_{\beta}>2$. To reduce the
conservative nature of these CVs, McCloskey (2017) suggests to use the
$1-\alpha+\eta$ quantile of $\mathcal{L}_{\infty}(\tilde{b},\tilde{b})$ for
some $\eta\in(0,\alpha)$, where $\tilde{b}$ is the maximizer across a
$(1-\eta)$-level confidence set for $b$. Unfortunately, this proposal suffers
a similar computational drawback to that of Andrews and Guggenberger (2009)
when $d_{\beta}>2$.

We finally note that the shrinkage-based bootstrap of Cavaliere, Nielsen,
Pedersen and Rahbek (2022) essentially seeks to choose between $\mathcal{L}%
_{\infty}(0,0)$ or $\mathcal{L}_{\infty}(\infty,\infty)$ (component-wise) in a
data-driven way. However, it fails to control size uniformly since there may
exist values $b\in(0,\infty)$ such that $\mathcal{L}_{\infty}(b,b)\geq
\mathcal{L}_{\infty}(0,0)$.

\subsection{Prerequisites and additional assumptions}

\label{sec ingredients}

In order to construct computationally-feasible and uniformly-valid CVs, we
require two main ingredients, which are very simple to satisfy in
applications. First, we need a consistent estimators of the covariance
matrices $\Omega_{0}$ and $\Sigma_{0}$, see Assumptions \ref{ass:hessian}%
--\ref{ass:score}. Second, we need the construction of a consistent and
asymptotically Gaussian estimator $\check{\beta}_{n}$ of $\beta_{n}$ with
asymptotic covariance matrix given by $\Sigma_{\beta}=H_{\beta}\Omega_{0}%
^{-1}\Sigma_{0}\Omega_{0}^{-1}H_{\beta}^{\prime}$ where $\Omega_{0}$ is
defined in Assumption \ref{ass:hessian} and $H_{\beta}$ is the selection
matrix satisfying $H_{\beta}\theta=\beta$. Moreover, we require a consistent
estimator $\check{\Sigma}_{\beta,n}$ of the covariance matrix $\Sigma_{\beta}%
$. We formalize these requirements through the following assumptions, which
hold for any sequence $\{\psi_{n}\}$ in $\Psi$ satisfying
(\ref{eq:def_drifting_sequence}).

\begin{assumption}
\label{ass:covariance-consistency} There exists a matrix $\hat{\Sigma}_{n}$
such that $\hat{\Sigma}_{n}\overset{p}{\rightarrow}\Sigma_{0}$.
\end{assumption}

\begin{assumption}
\label{ass:AN_estimator} There exists estimators $\check{\beta}_{n}$ and
$\check{\Sigma}_{\beta,n}$ which satisfy (i) $\sqrt{n}(\check{\beta}_{n}%
-\beta_{n})\overset{d}{\rightarrow}N(0,\Sigma_{\beta})$, with $\Sigma_{\beta
}=H_{\beta}\Omega_{0}^{-1}\Sigma_{0}\Omega_{0}^{-1}H_{\beta}^{\prime}$, and
(ii)$\ \check{\Sigma}_{\beta,n}\overset{p}{\rightarrow}\Sigma_{\beta}$.
\end{assumption}

As a simple example of an estimator satisfying Assumption
\ref{ass:AN_estimator} in general, consider the one-step iterated
Newton-Raphson estimator for $\theta$:
\begin{equation}
\check{\theta}_{n}=\hat{\theta}_{n}-\left(  \frac{\partial^{2}L_{n}%
(\hat{\theta}_{n})}{\partial\theta\partial\theta^{\prime}}\right)  ^{-1}%
\frac{\partial L_{n}(\hat{\theta}_{n})}{\partial\theta}.
\label{eq:NR_estimator}%
\end{equation}
As observed by Ketz (2018), by definition, $\check{\theta}_{n}$ may not belong
to the parameter space $\Theta$, and hence $L_{n}(\check{\theta}_{n})$ may not
even be well-defined. We have the following lemma due to Ketz (2018).

\begin{lemma}
\label{lem:Ketz} Under any sequence $\psi_{n}\in$ $\Psi$ satisfying
(\ref{eq:def_drifting_sequence}) and Assumptions \ref{ass:consistency}%
--\ref{ass:score}, with $\check{\theta}_{n}$ given by (\ref{eq:NR_estimator}%
),
\[
\sqrt{n}(\check{\theta}_{n}-\theta_{n})\overset{d}{\rightarrow}N(0,\Omega
_{0}^{-1}\Sigma_{0}\Omega_{0}^{-1}).
\]
In particular, we have that $\check{\beta}_{n}=H_{\beta}\check{\theta}_{n}$ is
asymptotically normal,
\begin{equation}
\sqrt{n}(\check{\beta}_{n}-\beta_{n})\overset{d}{\rightarrow}N(0,\Sigma
_{\beta}), \label{eq:AN_beta_bar}%
\end{equation}
where $\Sigma_{\beta}=H_{\beta}\Omega_{0}^{-1}\Sigma_{0}\Omega_{0}%
^{-1}H_{\beta}^{\prime}$. If, in addition, Assumption
\ref{ass:covariance-consistency} holds, then%
\begin{equation}
\check{\Sigma}_{\beta,n}=H_{\beta}\left(  n^{-1}\frac{\partial^{2}L_{n}%
(\hat{\theta}_{n})}{\partial\theta\partial\theta^{\prime}}\right)  ^{-1}%
\hat{\Sigma}_{n}\left(  n^{-1}\frac{\partial^{2}L_{n}(\hat{\theta}_{n}%
)}{\partial\theta\partial\theta^{\prime}}\right)  ^{-1}H_{\beta}^{\prime
}\overset{p}{\rightarrow}\Sigma_{\beta}\text{.} \label{eq:cons_Sigma_beta}%
\end{equation}

\end{lemma}

Hence, for the one-step iterated Newton-Raphson estimator $\check{\theta}_{n}$
of (\ref{eq:NR_estimator}) it holds that Assumptions \ref{ass:consistency}%
--\ref{ass:covariance-consistency} imply Assumption \ref{ass:AN_estimator}. We
illustrate this in terms of the two running examples.

\bigskip

\noindent\textsc{Running example: Regression. }To verify Assumptions
\ref{ass:covariance-consistency} and \ref{ass:AN_estimator} for the linear
regression example, we make the following additional assumptions.

\begin{liniidassumption}
\label{ass:linIID_moments}For all $\psi\in\Psi$ and some constants $\nu,c>0$,
$\mathbb{E}_{\psi}[|x_{i,t},x_{j,t},x_{k,t}x_{l,t}|^{1+\nu}]\leq c$ for
$i$,$j,k,l=1,\dots,d_{\theta}$.
\end{liniidassumption}

The following Proposition states that Assumptions
\ref{ass:covariance-consistency}--\ref{ass:AN_estimator} hold for the linear
regression example when choosing $\hat{\Sigma}_{n}$ as the Eicker-White
heteroskedasticity-robust estimator\footnote{Note that one could alternatively
define the estimator in terms of the the constrained residuals, that is,
$\hat{\varepsilon}_{t}=y_{t}-x_{t}^{\prime}\hat{\theta}_{n}$.},
\begin{equation}
\hat{\Sigma}_{n}=\frac{1}{n}\sum_{t=1}^{n}\hat{\varepsilon}_{t}^{2}x_{t}%
x_{t}^{\prime},\quad\hat{\varepsilon}_{t}=y_{t}-x_{t}^{\prime}\hat{\theta
}_{LS}. \label{eq:def_Eicker-White}%
\end{equation}

\begin{proposition}
\label{prop:reg ass verification 2} Suppose $\check{\beta}_{n}=\hat{\beta
}_{LS}$ and $\check{\Sigma}_{\beta,n}$ is equal to the upper $d_{\beta}\times
d_{\beta}$ block of $S_{xx}^{-1}\hat{\Sigma}_{n}S_{xx}^{-1}$ with $\hat
{\Sigma}_{n}$ given by (\ref{eq:def_Eicker-White}). Then Assumptions LinIID
\ref{ass:lin-iid}--\ref{ass:linIID_moments} imply Assumptions
\ref{ass:covariance-consistency}--\ref{ass:AN_estimator}.
\end{proposition}

\begin{remark}
\label{rem: time series regr}The linear regression example can easily be
extended to time series data. In particular, under appropriate strong mixing
conditions on $\{W_{t}\}_{t\in%
\mathbb{Z}
}$ and replacing $\hat{\Sigma}_{n}$ with a HAC estimator (Newey and West,
1987), it is possible to verify Assumptions \ref{ass:consistency}%
--\ref{ass:covariance-consistency}.\hfill$\square$
\end{remark}

\bigskip

\noindent\textsc{Running example: ARCH. }Under Assumptions ARCH
\ref{ass:ARCH-HL-stat-mixing}-\ref{ass:ARCH-HL-criterion} it holds that
$\Sigma_{0}=(\kappa/2)\Omega_{0}$; hence, we can estimate $\Sigma_{0}$ using
$\hat{\Sigma}_{n}=(\hat{\kappa}_{n}/2)\hat{\Omega}_{n}\ $with%
\[
\hat{\Omega}_{n}=-n^{-1}\frac{\partial^{2}L_{n}(\hat{\theta}_{n})}%
{\partial\theta\partial\theta^{\prime}},\text{ \hspace{0.77cm}}\hat{\kappa
}_{n}=n^{-1}\sum_{t=1}^{n}(\hat{\varepsilon}_{t}^{4}-1)
\]
where $\hat{\varepsilon}_{t}=y_{t}/\hat{\sigma}_{t}(\hat{\theta}_{n})$.
Moreover, let $\check{\beta}_{n}=H_{\beta}\check{\theta}_{n}$ with
$\check{\theta}_{n}$ given by (\ref{eq:NR_estimator}), and finally let
\begin{equation}
\check{\Sigma}_{\beta,n}=(\hat{\kappa}_{n}/2)H_{\beta}\hat{\Omega}_{n}%
^{-1}H_{\beta}^{\prime}. \label{eq:ARCH_Sigma_beta}%
\end{equation}

\begin{proposition}
\label{prop:ARCH ass verification 2}With $\check{\beta}_{n}$ and
$\check{\Sigma}_{\beta,n}$ as defined above, Assumptions ARCH
\ref{ass:ARCH-HL-stat-mixing}-\ref{ass:ARCH-HL-criterion} imply Assumptions
\ref{ass:covariance-consistency}--\ref{ass:AN_estimator}. \hfill$\square$
\end{proposition}

\subsection{Uniformly valid CV}

\label{sec our proposal}

We can now finally state our main algorithm to construct uniformly valid CVs,
which are denoted as \textsf{$CV$}$_{\alpha,n}$, where $\alpha\in(0,1)$
denotes the nominal significance level. As anticipated, it relies on the
asymptotically normal estimator $\check{\beta}_{n}$ and the consistent
covariance matrix estimator $\check{\Sigma}_{\beta,n}$ in Assumption
\ref{ass:AN_estimator}.

\begin{algorithm}
\label{algorithm}%
\mbox{} \vspace{0.3cm}%

With $\hat{\theta}_{n}$, $\hat{\Omega}_{n}=-n^{-1}\partial^{2}L_{n}%
(\hat{\theta}_{n})/\partial\theta\partial\theta^{\prime}$, $\hat{\Sigma}_{n}$,
$\check{\beta}_{n}$, $\check{\Sigma}_{\beta,n}$ and the nominal significance
level $\alpha\in\left(  0,1\right)  $ as inputs:

\begin{enumerate}
\item Choose some $\eta\in(0,\alpha)$;

\item Compute $\check{\Omega}_{\beta,n}=\operatorname*{diag}(\check{\Sigma
}_{\beta,n})^{-1/2}\check{\Sigma}_{\beta,n}\operatorname*{diag}(\check{\Sigma
}_{\beta,n})^{-1/2}$;

\item Compute $\check{q}_{1-\eta,n}$ as the $\left(  1-\eta\right)  $-quantile
of $\max_{i=1,\dots,d_{\beta}}|\check{Z}_{\beta,i}|$ with $\check{Z}_{\beta
}\overset{d}{=}N(0,\check{\Omega}_{\beta,n})$;

\item Compute $b_{L,n}=(b_{L,n,1},\ldots,b_{L,n,d_{\beta}})^{\prime}$ and
$b_{U,n}=(b_{U,n,1},\ldots,b_{U,n,d_{\beta}})^{\prime}$ by setting
$b_{L,n,i}=\max\{0,\check{b}_{L,n,i}\}$ and $b_{U,n,i}=\max\{0,\check
{b}_{U,n,i}\}$ for $i=1,\ldots,d_{\beta}$, where
\[
\check{b}_{L,n}=\sqrt{n}\check{\beta}_{n}-\check{q}_{1-\eta,n}%
\operatorname*{diagv}(\check{\Sigma}_{\beta,n})^{1/2}\text{, }\check{b}%
_{U,n}=\sqrt{n}\check{\beta}_{n}+\check{q}_{1-\eta,n}\operatorname*{diagv}%
(\check{\Sigma}_{\beta,n})^{1/2}\text{;}%
\]

\item Compute $\mathsf{CV}_{\alpha,n}$ as the $1-\alpha+\eta$ quantile of
$\mathcal{L}_{\infty,n}(b_{L,n},b_{U,n})$, where $\mathcal{L}_{\infty,n}%
(\cdot,\cdot)$ is defined as $\mathcal{L}_{\infty}(\cdot,\cdot)$ in
(\ref{eq:def:L_infty}) with $\Omega_{0}$ and $\Sigma_{0}$ replaced by
$\check{\Omega}_{n}$ and $\hat{\Sigma}_{n}$, respectively.
\end{enumerate}
\end{algorithm}

Notice that the quantiles in Steps 3 and 5 of Algorithm \ref{algorithm} can be
computed straightforwardly by simulation. Notice also that the elements of
$b_{L,n}$ and $b_{U,n}$ are maximized against zero in order to ensure that
$b_{L,n}$ and $b_{U,n}$ obeys the lower bounds of $b$.

We can now state our main theorem, where we establish the uniform asymptotic
size control of the LR test based on our proposed critical value,
\textsf{$CV$}$_{\alpha,n}$.

\begin{theorem}
\label{thm:asysize control} Let $\mathsf{CV}_{\alpha,n}$ be constructed as in
Algorithm \ref{algorithm}. Then under Assumptions \ref{ass:consistency}%
--\ref{ass:AN_estimator},%
\[
\limsup_{n\rightarrow\infty}\sup_{\psi\in\Psi}\mathbb{P}_{\psi}\left(
\mathsf{LR}_{n}\geq\mathsf{CV}_{\alpha,n}\right)  \leq\alpha\text{.}%
\]

\end{theorem}

We emphasize that the underlying assumptions of Theorem
\ref{thm:asysize control} are intuitive and typically possible to very for a
given statistical model. In particular, have that the critical values
determined by Algorithm \ref{algorithm} are valid in terms of the linear
regression and the ARCH models due to Propositions
\ref{prop:reg ass verification 2} and \ref{prop:ARCH ass verification 2},
respectively. The result states that Algorithm \ref{algorithm} provides
asymptotically valid CVs for any choice of $\eta\in(0,\alpha)$. We note that
$1-\eta$ (approximately) quantifies the probability of the event $b\in\lbrack
b_{L,n},b_{U,n}]$. The parameter $\eta$ is user-chosen and following the
existing body of literature (e.g., McCloskey, 2017) we recommend choosing
$\eta=\alpha/10$.

\section{Simulation experiments\label{sec:sim}}

In this section, we consider the performance of our proposed LR test. Section
\ref{sec:sim_lin} considers the linear regression with positivity-constrained
coefficients as in the running example, whereas Section \ref{sec:sim_ARCH}
considers the ARCH model.

\subsection{Linear regression with positivity constraints\label{sec:sim_lin}}

Consider the linear regression model, given by%
\[
y_{t}=\gamma x_{t,1}+\beta x_{t,2}+\varepsilon_{t},\quad t=1,\dots,n,
\]
where for $x_{t}=(x_{t,1},x_{t,2})^{\prime}$, $\{(x_{t}^{^{\prime}%
},\varepsilon_{t})^{\prime}\}_{t=1,2,\dots}$ an i.i.d. process with%
\[%
\begin{pmatrix}
x_{t}\\
\varepsilon_{t}%
\end{pmatrix}
\sim N\left(  0,%
\begin{pmatrix}
\Omega & 0\\
0 & 1
\end{pmatrix}
\right)  ,
\]
and $\theta=(\gamma,\beta^{\prime})^{\prime}$ with $\gamma,\beta\in
\lbrack0,\infty)$. The matrix $\Omega$ is a correlation matrix given by
\[
\Omega=\left(
\begin{array}
[c]{cc}%
1 & \rho\\
\rho & 1
\end{array}
\right)  .
\]
We seek to test the hypothesis $\mathsf{H}_{0}:\gamma=0$ against $\gamma>0$.

For comparison, we report the rejection frequencies for the LR\ test where one
ignores that $\beta$ is constrained. Here the CV is given by 2.71, the 90th
percentile of the $\chi_{1}^{2}$ distribution, which is motivated by
(erroneously) approximating the distribution of the LR statistic by its
asymptotic distribution at the boundary point $\beta=0$, $(\max
\{0,N(0,1)\})^{2}$ (labelled \textquotedblleft LR\textquotedblright\ in the
tables). We also compare with the conditional LR (\textquotedblleft
CLR\textquotedblright\ in the tables) test proposed by Ketz (2018). All tests
are carried out at a nominal level of $\alpha=5\%$, and for the uniform CV
construction $\eta=\alpha/10$. For the uniform CV construction as well as the
CLR test, the CVs are determined by means of simulation, making use of 10,000
draws. All rejection frequencies are based on 10,000 Monte Carlo replications.

Tables 1-4 report the rejection frequencies of our proposed test for different
values of $\gamma,$ $\beta,$ $\rho,$ and $n$. Specifically, we consider the
cases $\gamma,\beta\in\{0,0.1\}$, $\rho\in\{-0.95,-0.75,0.5,0,0.5,0.75,0.95\}$
and sample sizes $n\in\{100,250,500,1000\}.$

Tables 1 and 2 contain the rejection frequencies under $\mathsf{H}_{0}$.

\begin{center}
[Tables 1 and 2 around here]
\end{center}

Our test appears to control size. It tends to be conservative whenever
$(\max(N(0,1),0)^{2}$ yields conservative CVs, that is for correlations
$\rho\geq0.5$. Importantly, it controls size when $(\max(N(0,1),0)^{2}$ does
not, that is for correlations $\rho\leq-0.5$. The CLR test has rejection
frequencies around $\alpha$ across all correlations.

Tables 3 and 4 contains the rejection frequencies under the alternative.

\begin{center}
[Tables 3 and 4 around here]
\end{center}

Our proposed method appears to have attractive rejection frequencies under
most alternatives, and in particular it has higher rejection frequencies than
the CLR approach for correlations $\rho\leq-0.75$.

\subsection{ARCH\label{sec:sim_ARCH}}

In this section we consider the ARCH model and seek to evaluate the
performance of our proposed method when testing for the presence of an
explanatory covariate when another covariate may be present. Specifically,
consider the model%
\begin{align*}
y_{t}  &  =\sigma_{t}\varepsilon_{t},\quad t=1,\dots,n,\\
\sigma_{t}^{2}  &  =\delta_{1}+\delta_{2}y_{t-1}^{2}+\gamma x_{t-1,1}+\beta
x_{t-1,2},
\end{align*}
where $\delta_{1}>0$, $\delta_{2},\gamma,\beta\geq0$, and $\{\varepsilon
_{t}\}_{t=1,\dots,n}$ is an i.i.d. process with $\varepsilon_{t}\sim N(0,1)$.
We consider testing $\mathsf{H}_{0}:\gamma=0$, against $\gamma>0$.

In terms of the covariates, we follow the simulation design in Nielsen,
Pedersen, Rahbek and Thorsen (2024) and let $(x_{t,1},x_{t,2})=(\exp
(v_{t,1}),\exp(v_{t,2}))$ with $V_{t}=\left(  v_{1,t},v_{2,t}\right)
^{\prime}$ a bivariate autoregression with correlated innovations satisfying
$V_{t}=aV_{t-1}+\epsilon_{t}$, $t=1,\dots,n$. Here $\left\{  \epsilon
_{t}\right\}  _{t=1}^{n}$ is an i.i.d. $N_{2}(0,\Sigma)$ process with%
\[
\Sigma=b\left(  1-a^{2}\right)  \left(
\begin{array}
[c]{cc}%
1 & \rho_{12}\\
\rho_{12} & 1
\end{array}
\right)
\]
independent of $\left\{  z_{t}\right\}  _{t=1}^{n},$ $\rho_{12}\in(-1,1),$
$a=0.9$ and $b=0.5$. The simulated realizations of the processes for $y_{t}$
and $(x_{t,1},x_{t,2})$ make use of a burn-in period of 1000 observations.

For the experiment we assume that it is known to the researcher that the true
value of the ARCH coefficient $\delta_{2}$ is not near its boundary of zero,
so that the only parameter that potentially causes a discontinuity in the null
distribution is $\beta$. We report the rejection frequencies for $n=1000$
observations, parameter values $\gamma,\beta\in\{0,0.01,0.05,0.1,0.25\}$,
$\rho_{12}\in\{-0.95,-0.75,0.5,0,0.5,0.75,0.95\}$, and compare with the
standard LR with CVs derived at the boundary as well as the CLR test as in the
previous section.

Table 5 contains rejection frequencies under $\mathsf{H}_{0}$ for different
values of $\beta$ and $\rho_{12}$. We note that similar to the findings for
the linear regression case in the previous section, our proposed method has
rejection frequencies less than $\alpha$ for all combinations of parameter
values. Similar to the linear regression case, the standard LR test
overrejects for the case of $\beta=0$ and small values of $\rho_{12}$.

\begin{center}
[Tables 5 and 6 around here]
\end{center}

Table 6 contains the rejection frequencies for the alternatives $\gamma
\in\{0.01,0.05,0.1,0.25\}$ with $\beta=0$. Our proposed method performs well
in terms of rejecting $\mathsf{H}_{0}$, and performs comparably to the
CLR\ test for most combinations of the parameter values.

Summarizing the findings in Sections \ref{sec:sim_lin} and \ref{sec:sim_ARCH},
we have that our proposed method yields attractive rejection frequencies in
most settings and yield more powerful tests than the CLR approach in cases
with extreme correlations between covariates.

\section{Empirical illustration\label{sec:emp_illustration}}

In this section we consider an empirical illustration of our proposed test. We
consider ARCH models for daily returns of various stock indices and test for
the presence of ARCH effects and spillovers from the U.S. stock market.
Inspired by the HAR model of Corsi (2009), and similar to Nielsen et al.
(2024, Section 5), we include lagged Realized Volatility (RV) covariates to
account for potential high persistence in the conditional variance of the
index returns. With the \emph{daily} index return given by $y_{t}$ in
(\ref{eq:ARCHX_y}), let
\[
\sigma_{t}^{2}=\delta+\beta_{ARCH}y_{t-1}^{2}+\beta_{RV}RV_{t-1}+\beta
_{W}RV_{W,t-1}+\beta_{M}RV_{M,t-1}+\beta_{SPX}SPX_{t-1}^{2},
\]
with $\delta>0$ and $\beta_{ARCH},\beta_{RV},\beta_{M},\beta_{SPX}\geq0$.\ The
variable $RV_{t}$ is the RV of the index based on 5-minutes intraday returns
at day $t$, $RV_{W,t}=5^{-1}\sum_{i=0}^{4}RV_{t-i}$ is the weekly average RV,
and $RV_{W,t}=22^{-1}\sum_{i=0}^{21}RV_{t-i}$ is the monthly average RV.
Lastly, the variable $SPX_{t}$ is the (continuously compounded,
close-to-close) return on the S\&P 500 index at day $t$. We refer to Nielsen
et al. (2024, Section 5) for a discussion and motivation for this type of
model. Based on data retrieved from the Oxford Man Realized Library covering
the period 3 January 2000 to 27 June 2018, we analyze the following
indices:\ Australian ASX All Ordinaries (AORD), Belgian BEL 20 (BFX), Spanish
IBEX 35 (IBEX), IPC\ Mexico (MXX), Indian NIFTY50 (NSEI) and Danish OMX C20
(OMX) (notice that OXM\ C20 begins on 3 October 2005). For each of these
indices, we test the hypothesis for no ARCH, $\mathsf{H}_{ARCH}:\beta
_{ARCH}=0$, and the hypothesis of no spillovers from the U.S. stock market,
$\mathsf{H}_{SPX}:\beta_{SPX}=0$. The tests are carried out under the
assumption that the true value $\delta_{0}$ is away from its lower bound,
$\delta_{0}>\delta_{L}$, but no assumptions are imposed on the remaining
nuisance parameters.

Table 7 contains point estimates of the model parameters for each index
series. Moreover, it contains the values of the LR and CLR\ statistics along
with CVs based on 5\% nominal levels.\ Based on the LR test, for all of the
series except for the Indian NSEI index, we cannot reject $\mathsf{H}_{ARCH}$.
On the contrary we reject $\mathsf{H}_{SPX}$ for all series except for the
Danish OMX. Note that if one uses as CV, 2.71, based on the $\max
\{N(0,1),0\}^{2}$-distribution, one would reject $\mathsf{H}_{SPX}$ also for
the OMX series.

\begin{center}
[Table 7 around here]
\end{center}

In short, based on our proposed critical values, we find evidence for no ARCH
effects in most of the index return series, whereas most of the series appear
subject to spillovers from the U.S. stock market.

\section{Conclusions}

This paper proposes a novel and computationally efficient method for
constructing CVs for LR tests that achieve uniform asymptotic size control,
even when nuisance parameters lie on or near the boundary of their parameter
space. The key innovation lies in using confidence bounds for an
asymptotically Gaussian approximation of a nuisance parameter estimator and
exploiting the monotonicity properties of the two components of the LR
statistic's asymptotic distribution. This allows for the formation of valid
CVs via straightforward Monte Carlo simulation, without the need for intensive
optimization or conservative least-favorable approaches. The method remains
tractable in settings for which the nuisance parameter is not low-dimensional
and is broadly applicable, including to models such as constrained regressions
and ARCH specifications, offering the first uniformly valid test for the latter.

\section*{References}

\begin{description}
\item \textbf{\smallskip\noindent}\textsc{Andrews, D.W.K.} (2001):
\textquotedblleft Testing when a parameter is on the boundary of the
maintained hypothesis\textquotedblright, \emph{Econometrica}, vol. 69, 683--734.

\item \textbf{\smallskip\noindent}\textsc{Andrews, D.W.K.} \textsc{and Cheng,
X. }(2012),\textquotedblleft Estimation and inference with weak, semi-strong,
and strong identification\textquotedblright, \emph{Econometrica}, vol. 80, 2153--2211.

\item \textbf{\smallskip\noindent}\textsc{Andrews, D.W.K.} \textsc{and
Guggenberger, P}. (2009): \textquotedblleft Hybrid and size-corrected
subsampling methods\textquotedblright, \emph{Econometrica}, vol. 77, 721--762.

\item \textbf{\smallskip\noindent}\textsc{Berger, R. L. and Boos, D. D.}
(1994): \textquotedblleft P values maximized over a confidence set for the
nuisance parameter\textquotedblright, \emph{Journal of the American
Statistical Association}, vol. 89, 1012--1016.

\item \textbf{\smallskip\noindent}\textsc{Berry, S., Levinsohn, J., and Pakes,
A.} (1995): \textquotedblleft Automobile prices in market
equilibrium\textquotedblright, \emph{Econometrica}, vol. 63, 841--890.

\item \textbf{\smallskip\noindent}\textsc{Cavaliere, G., Nielsen, H. B.,
Pedersen, R. S., and Rahbek, A}. (2022): \textquotedblleft Bootstrap inference
on the boundary of the parameter space, with application to conditional
volatility models\textquotedblright, \emph{Journal of Econometrics}, vol. 227, 241--263.

\item \textbf{\smallskip\noindent}\textsc{Cavaliere, G., Nielsen, H. B., and
Rahbek, A}. (2017): \textquotedblleft On the consistency of bootstrap testing
for a parameter on the boundary of the parameter space\textquotedblright,
\emph{Journal of Time Series Analysis}, vol. 38, 513--534.

\item \textbf{\smallskip\noindent}\textsc{Cavaliere, G., Perera, I., and
Rahbek, A.} (2024): \textquotedblleft Specification tests for GARCH
processes\textquotedblright, \emph{Journal of Business and Economic
Statistics}, vol. 42, 197--214.

\item \textbf{\smallskip\noindent}\textsc{Corsi, F.} (2009): \textquotedblleft
A simple approximate long-memory model of realized
volatility\textquotedblright, \emph{Journal of Financial Econometrics}, vol.
7, 174--196.

\item \textbf{\smallskip\noindent}\textsc{Drton, M.} (2009): \textquotedblleft
Likelihood ratio tests and singularities\textquotedblright, \emph{Annals of
Statistics}, vol. 37, 979--1012.

\item \textbf{\smallskip\noindent}\textsc{Fan, Y. and Shi, X.} (2023):
\textquotedblleft Wald, QLR, and score tests when the parameters are subject
to linear inequality constraints\textquotedblright, \emph{Journal of
Econometrics}, vol. 235, 2005-2026.

\item \textbf{\smallskip\noindent}\textsc{Francq, C. and Thieu, L. Q.} (2019):
\textquotedblleft QML inference for volatility models with
covariates\textquotedblright, \emph{Econometric Theory}, vol. 35, 37--72.

\item \textbf{\smallskip\noindent}\textsc{Francq, C., and Zako\"{\i}an, J.-M}.
(2009), \textquotedblright Testing the nullity of GARCH coefficients:
Correction of the standard tests and relative efficiency
comparisons\textquotedblleft, \emph{Journal of the American Statistical
Association}, vol. 104, 313--324.

\item \textbf{\smallskip\noindent}\textsc{Geyer, C.\ J}.
(1994),\textquotedblleft On the asymptotics of constrained $M$%
-estimators\textquotedblright, \emph{The Annals of Statistics}, vol. 22, 1993--2010.

\item \textbf{\smallskip\noindent}\textsc{Hong, H. and Li, J.}
(2020),\textquotedblleft The numerical bootstrap\textquotedblright, \emph{The
Annals of Statistics}, vol. 48, 397--412.

\item \textbf{\smallskip\noindent}\textsc{Ketz, P}. (2018): \textquotedblleft
Subvector inference when the true parameter vector may be near or at the
boundary\textquotedblright, \emph{Journal of Econometrics}, vol. 207, 285--306.

\item \textbf{\smallskip\noindent}\textsc{Ketz, P. and McCloskey, A}. (2023):
\textquotedblleft Short and simple confidence intervals when the directions of
some effects are known\textquotedblright, \emph{Review of Economics and
Statistics}, forthcoming.

\item \textbf{\smallskip\noindent}\textsc{Li, J.} (2025),\textquotedblleft The
proximal bootstrap for constrained estimators\textquotedblright, \emph{Journal
of Statistical Planning and Inference}, vol. 236, 106245.

\item \textbf{\smallskip\noindent}\textsc{McCloskey, A. }(2017):
\textquotedblleft Bonferroni-based size-correction for nonstandard testing
problems\textquotedblright, \emph{Journal of Econometrics}, vol. 200, 17--35.

\item \textbf{\smallskip\noindent}\textsc{Mitchell, J. D., Allman, E. S., and
Rhodes, J. A}. (2019): \textquotedblleft Hypothesis testing near singularities
and boundaries\textquotedblright, \emph{Electronic Journal of Statistics},
vol. 13, 2150--2193.

\item \textbf{\smallskip\noindent}\textsc{Newey, W. K. and McFadden, D. L}.
(1994): \textquotedblleft Large sample estimation and hypothesis testing". In
Engle, R. and McFadden, D., editors, \emph{Handbook of Econometrics}, vol. 4, 2111--2245.

\item \textbf{\smallskip\noindent}\textsc{Newey, W. K. and West, K. D.}
(1987): \textquotedblleft A simple, positive semi-definite, heteroskedasticity
and autocorrelation consistent covariance matrix\textquotedblright,
\emph{Econometrica}, vol. 55, 703--708.

\item \textbf{\smallskip\noindent}\textsc{Nielsen, H. B., Pedersen, R. S.,
Rahbek, A., and Thorsen,} S. N.,\textquotedblleft Testing in GARCH-X\ models:
Boundary, correlations and bootstrap theory\textquotedblright, \emph{Journal
of Time Series Analysis}, forthcoming.

\item \textbf{\smallskip\noindent}\textsc{Self, S. G. and Liang, K.-Y}.
(1987),\textquotedblleft Asymptotic properties of maximum likelihood
estimators and likelihood ratio tests under nonstandard
conditions\textquotedblright, \emph{Journal of the American Statistical
Association}, vol. 82, 605--610.

\item \textbf{\smallskip\noindent}\textsc{Shapiro}, A.
(1989),\textquotedblleft Asymptotic properties of statistical estimators in
stochastic programming\textquotedblright, \emph{The Annals of Statistics},
vol. 17, 841--858.

\item \textbf{\smallskip\noindent}\textsc{Silvapulle, M.\ J.}
(1996),\textquotedblleft A test in the presence of nuisance
parameters\textquotedblright, \emph{Journal of the American Statistical
Association}, vol. 91, 1690--1693.

\item \textbf{\smallskip\noindent}\textsc{Silvapulle, M.\ J. and Silvapulle,
P.} (1995),\textquotedblleft A score test against one-sided
alternatives\textquotedblright, \emph{Journal of the American Statistical
Association}, vol. 90, 342--349.
\end{description}

\newpage

%

\thispagestyle{empty}%
%

\begin{table}[h]\label{tab1}
\centering\caption{Rejection Frequencies under null hypothesis, $ \protect
(\gamma, \beta) = (0,0) $}
\small\begin{tabular}{c|ccc}
\toprule$n$ & LR & CLR & LR-uniform \\
\midrule\multicolumn{4}{c}{$\rho= -0.95$} \\
100 & 0.0998 & 0.0520 & 0.0414 \\
250 & 0.1050 & 0.0526 & 0.0397 \\
500 & 0.1054 & 0.0536 & 0.0422 \\
1000 & 0.1040 & 0.0502 & 0.0424 \\
\midrule\multicolumn{4}{c}{$\rho= -0.75$} \\
100 & 0.0813 & 0.0527 & 0.0352 \\
250 & 0.0873 & 0.0545 & 0.0359 \\
500 & 0.0820 & 0.0513 & 0.0332 \\
1000 & 0.0867 & 0.0529 & 0.0367 \\
\midrule\multicolumn{4}{c}{$\rho= -0.5$} \\
100 & 0.0727 & 0.0523 & 0.0299 \\
250 & 0.0684 & 0.0505 & 0.0297 \\
500 & 0.0753 & 0.0528 & 0.0309 \\
1000 & 0.0677 & 0.0482 & 0.0250 \\
\midrule\multicolumn{4}{c}{$\rho= 0$} \\
100 & 0.0520 & 0.0545 & 0.0223 \\
250 & 0.0489 & 0.0512 & 0.0183 \\
500 & 0.0510 & 0.0513 & 0.0191 \\
1000 & 0.0482 & 0.0498 & 0.0190 \\
\midrule\multicolumn{4}{c}{$\rho= 0.5$} \\
100 & 0.0293 & 0.0547 & 0.0114 \\
250 & 0.0283 & 0.0501 & 0.0105 \\
500 & 0.0288 & 0.0512 & 0.0099 \\
1000 & 0.0286 & 0.0519 & 0.0103 \\
\midrule\multicolumn{4}{c}{$\rho= 0.75$} \\
100 & 0.0142 & 0.0552 & 0.0049 \\
250 & 0.0143 & 0.0584 & 0.0041 \\
500 & 0.0138 & 0.0550 & 0.0043 \\
1000 & 0.0152 & 0.0577 & 0.0046 \\
\midrule\multicolumn{4}{c}{$\rho= 0.95$} \\
100 & 0.0014 & 0.0602 & 0.0011 \\
250 & 0.0014 & 0.0577 & 0.0010 \\
500 & 0.0015 & 0.0634 & 0.0011 \\
1000 & 0.0013 & 0.0603 & 0.0010 \\
\midrule\bottomrule\end{tabular}
\end{table}%

\newpage%

\thispagestyle{empty}%
%

\begin{table}[h]
\centering\caption{Rejection Frequencies under null hypothesis, $ \protect
(\gamma, \beta) = (0,0.1) $}
\small\begin{tabular}{c|ccc}
\toprule$n$ & LR & CLR & LR-uniform \\
\midrule\multicolumn{4}{c}{$\rho= -0.95$} \\
100 & 0.0562 & 0.0537 & 0.0224 \\
250 & 0.0540 & 0.0536 & 0.0205 \\
500 & 0.0495 & 0.0493 & 0.0217 \\
1000 & 0.0488 & 0.0507 & 0.0237 \\
\midrule\multicolumn{4}{c}{$\rho= -0.75$} \\
100 & 0.0560 & 0.0553 & 0.0226 \\
250 & 0.0527 & 0.0530 & 0.0235 \\
500 & 0.0547 & 0.0554 & 0.0294 \\
1000 & 0.0510 & 0.0507 & 0.0369 \\
\midrule\multicolumn{4}{c}{$\rho= -0.5$} \\
100 & 0.0554 & 0.0544 & 0.0226 \\
250 & 0.0514 & 0.0519 & 0.0250 \\
500 & 0.0470 & 0.0468 & 0.0257 \\
1000 & 0.0488 & 0.0487 & 0.0339 \\
\midrule\multicolumn{4}{c}{$\rho= 0$} \\
100 & 0.0522 & 0.0543 & 0.0213 \\
250 & 0.0505 & 0.0512 & 0.0235 \\
500 & 0.0497 & 0.0514 & 0.0247 \\
1000 & 0.0526 & 0.0523 & 0.0302 \\
\midrule\multicolumn{4}{c}{$\rho= 0.5$} \\
100 & 0.0423 & 0.0538 & 0.0155 \\
250 & 0.0499 & 0.0544 & 0.0195 \\
500 & 0.0493 & 0.0504 & 0.0170 \\
1000 & 0.0460 & 0.0463 & 0.0169 \\
\midrule\multicolumn{4}{c}{$\rho= 0.75$} \\
100 & 0.0288 & 0.0543 & 0.0083 \\
250 & 0.0367 & 0.0487 & 0.0113 \\
500 & 0.0445 & 0.0502 & 0.0148 \\
1000 & 0.0496 & 0.0510 & 0.0185 \\
\midrule\multicolumn{4}{c}{$\rho= 0.95$} \\
100 & 0.0048 & 0.0526 & 0.0017 \\
250 & 0.0108 & 0.0536 & 0.0028 \\
500 & 0.0226 & 0.0547 & 0.0065 \\
1000 & 0.0364 & 0.0543 & 0.0085 \\
\midrule\bottomrule\end{tabular}
\end{table}%

\newpage%

\thispagestyle{empty}%
%

\begin{table}[h]
\centering\caption
{Rejection Frequencies under alternative hypothesis, $ \protect(\gamma
, \beta) = (0.1,0) $}
\small\begin{tabular}{c|ccc}
\toprule$n$ & LR & CLR & LR-uniform \\
\midrule\multicolumn{4}{c}{$\rho= -0.95$} \\
100 & 0.3214 & 0.0947 & 0.1811 \\
250 & 0.5331 & 0.1250 & 0.3518 \\
500 & 0.7741 & 0.1780 & 0.6037 \\
1000 & 0.9504 & 0.2607 & 0.8811 \\
\midrule\multicolumn{4}{c}{$\rho= -0.75$} \\
100 & 0.3232 & 0.1707 & 0.1870 \\
250 & 0.5326 & 0.2726 & 0.3529 \\
500 & 0.7638 & 0.4411 & 0.6022 \\
1000 & 0.9432 & 0.6702 & 0.8702 \\
\midrule\multicolumn{4}{c}{$\rho= -0.5$} \\
100 & 0.2947 & 0.2150 & 0.1676 \\
250 & 0.5257 & 0.3929 & 0.3487 \\
500 & 0.7499 & 0.6147 & 0.5870 \\
1000 & 0.9443 & 0.8611 & 0.8758 \\
\midrule\multicolumn{4}{c}{$\rho= 0$} \\
100 & 0.2665 & 0.2708 & 0.1508 \\
250 & 0.4712 & 0.4774 & 0.3061 \\
500 & 0.7124 & 0.7146 & 0.5519 \\
1000 & 0.9302 & 0.9305 & 0.8542 \\
\midrule\multicolumn{4}{c}{$\rho= 0.5$} \\
100 & 0.1720 & 0.2246 & 0.0902 \\
250 & 0.3474 & 0.3961 & 0.1997 \\
500 & 0.5695 & 0.6088 & 0.3966 \\
1000 & 0.8498 & 0.8599 & 0.7174 \\
\midrule\multicolumn{4}{c}{$\rho= 0.75$} \\
100 & 0.0958 & 0.1660 & 0.0432 \\
250 & 0.2108 & 0.2783 & 0.1028 \\
500 & 0.3851 & 0.4332 & 0.2238 \\
1000 & 0.6554 & 0.6733 & 0.4730 \\
\midrule\multicolumn{4}{c}{$\rho= 0.95$} \\
100 & 0.0106 & 0.0921 & 0.0036 \\
250 & 0.0290 & 0.1232 & 0.0105 \\
500 & 0.0752 & 0.1737 & 0.0237 \\
1000 & 0.1796 & 0.2560 & 0.0620 \\
\midrule\bottomrule\end{tabular}
\end{table}%

\newpage%

\thispagestyle{empty}%
%

\begin{table}[h]
\centering\caption
{Rejection Frequencies under alternative hypothesis $ \protect(\gamma
, \beta) = (0.1,0.1) $}
\small\begin{tabular}{c|ccc}
\toprule$n$ & LR & CLR & LR-uniform \\
\midrule\multicolumn{4}{c}{$\rho= -0.95$} \\
100 & 0.1474 & 0.0927 & 0.0691 \\
250 & 0.1804 & 0.1229 & 0.0803 \\
500 & 0.2267 & 0.1716 & 0.1114 \\
1000 & 0.3168 & 0.2598 & 0.1666 \\
\midrule\multicolumn{4}{c}{$\rho= -0.75$} \\
100 & 0.1991 & 0.1690 & 0.1010 \\
250 & 0.3152 & 0.2853 & 0.1820 \\
500 & 0.4570 & 0.4335 & 0.2931 \\
1000 & 0.6855 & 0.6729 & 0.5117 \\
\midrule\multicolumn{4}{c}{$\rho= -0.5$} \\
100 & 0.2294 & 0.2190 & 0.1251 \\
250 & 0.4054 & 0.3947 & 0.2557 \\
500 & 0.6235 & 0.6196 & 0.4553 \\
1000 & 0.8668 & 0.8649 & 0.7586 \\
\midrule\multicolumn{4}{c}{$\rho= 0$} \\
100 & 0.2588 & 0.2659 & 0.1466 \\
250 & 0.4656 & 0.4719 & 0.3103 \\
500 & 0.7107 & 0.7118 & 0.5699 \\
1000 & 0.9313 & 0.9314 & 0.8887 \\
\midrule\multicolumn{4}{c}{$\rho= 0.5$} \\
100 & 0.2071 & 0.2269 & 0.1044 \\
250 & 0.3861 & 0.3921 & 0.2311 \\
500 & 0.6121 & 0.6135 & 0.4495 \\
1000 & 0.8640 & 0.8639 & 0.7921 \\
\midrule\multicolumn{4}{c}{$\rho= 0.75$} \\
100 & 0.1416 & 0.1700 & 0.0630 \\
250 & 0.2724 & 0.2818 & 0.1427 \\
500 & 0.4353 & 0.4369 & 0.2721 \\
1000 & 0.6657 & 0.6660 & 0.5083 \\
\midrule\multicolumn{4}{c}{$\rho= 0.95$} \\
100 & 0.0305 & 0.0964 & 0.0099 \\
250 & 0.0822 & 0.1269 & 0.0255 \\
500 & 0.1567 & 0.1712 & 0.0623 \\
1000 & 0.2527 & 0.2527 & 0.1272 \\
\midrule\bottomrule\end{tabular}
\end{table}%

\newpage%

\thispagestyle{empty}%
\begin{table}[h]
\centering\small\caption{Rejection Frequencies under null hypothesis}
\begin{tabular}{c|ccc}
\toprule$\beta_2$ & LR & CLR & LR-Uniform \\
\midrule\multicolumn{4}{c}{$\rho_{12} = -0.95$} \\
0 & 0.07814 & 0.03445 & 0.03094 \\
0.01 & 0.05682 & 0.04196 & 0.0257 \\
0.05 & 0.0504 & 0.03612 & 0.04296 \\
0.1 & 0.05299 & 0.03883 & 0.05491 \\
0.25 & 0.04597 & 0.04638 & 0.05346 \\
\midrule\multicolumn{4}{c}{$\rho_{12} = -0.75$} \\
0 & 0.07296 & 0.03733 & 0.03252 \\
0.01 & 0.04922 & 0.0389 & 0.02226 \\
0.05 & 0.05032 & 0.03988 & 0.04359 \\
0.1 & 0.05171 & 0.04084 & 0.05311 \\
0.25 & 0.05149 & 0.04218 & 0.05462 \\
\midrule\multicolumn{4}{c}{$\rho_{12} = -0.5$} \\
0 & 0.06602 & 0.03927 & 0.02835 \\
0.01 & 0.05276 & 0.04005 & 0.02233 \\
0.05 & 0.05114 & 0.03901 & 0.04231 \\
0.1 & 0.04819 & 0.03945 & 0.0503 \\
0.25 & 0.04596 & 0.03941 & 0.0504 \\
\midrule\multicolumn{4}{c}{$\rho_{12} = 0$} \\
0 & 0.04871 & 0.03818 & 0.01854 \\
0.01 & 0.04683 & 0.03913 & 0.02071 \\
0.05 & 0.05045 & 0.03964 & 0.03844 \\
0.1 & 0.05148 & 0.04187 & 0.05158 \\
0.25 & 0.04842 & 0.03589 & 0.04882 \\
\midrule\multicolumn{4}{c}{$\rho_{12} = 0.5$} \\
0 & 0.02877 & 0.05214 & 0.01233 \\
0.01 & 0.04805 & 0.04324 & 0.01992 \\
0.05 & 0.04851 & 0.04021 & 0.0245 \\
0.1 & 0.04613 & 0.03792 & 0.03592 \\
0.25 & 0.04931 & 0.03981 & 0.04771 \\
\midrule\multicolumn{4}{c}{$\rho_{12} = 0.75$} \\
0 & 0.01475 & 0.0620 & 0.00612 \\
0.01 & 0.03772 & 0.04913 & 0.01501 \\
0.05 & 0.0488 & 0.0434 & 0.0189 \\
0.1 & 0.0504 & 0.0454 & 0.0230 \\
0.25 & 0.0482 & 0.0430 & 0.0385 \\
\midrule\multicolumn{4}{c}{$\rho_{12} = 0.95$} \\
0 & 0.0009022 & 0.05704 & 0.001403 \\
0.01 & 0.0106 & 0.05932 & 0.003701 \\
0.05 & 0.0421 & 0.0489 & 0.0118 \\
0.1 & 0.0538 & 0.0537 & 0.0217 \\
0.25 & 0.0481 & 0.0481 & 0.0188 \\
\midrule\bottomrule\end{tabular}
\end{table}%

\newpage%

\thispagestyle{empty}%
\begin{table}[h]
\centering\small\caption{Rejection Frequencies for under various alternatives}
\begin{tabular}{c|ccc}
\toprule$\gamma$ & LR & CLR & LR-uniform \\
\midrule\multicolumn{4}{c}{$\rho_{12} = -0.95$} \\
0 & 0.08134 & 0.0379 & 0.03298 \\
0.01 & 0.4380 & 0.2167 & 0.2813 \\
0.05 & 0.9902 & 0.7901 & 0.9649 \\
0.1 & 0.9998 & 0.9114 & 0.9965 \\
0.25 & 1.0000 & 0.9568 & 0.9981 \\
\midrule\multicolumn{4}{c}{$\rho_{12} = -0.75$} \\
0 & 0.0701 & 0.03691 & 0.02888 \\
0.01 & 0.4377 & 0.2610 & 0.2872 \\
0.05 & 0.9891 & 0.8976 & 0.9687 \\
0.1 & 1.0000 & 0.9714 & 0.9982 \\
0.25 & 1.0000 & 0.9856 & 0.9986 \\
\midrule\multicolumn{4}{c}{$\rho_{12} = -0.5$} \\
0 & 0.06329 & 0.03815 & 0.02544 \\
0.01 & 0.4230 & 0.2956 & 0.2765 \\
0.05 & 0.9889 & 0.9527 & 0.9688 \\
0.1 & 0.9999 & 0.9939 & 0.9988 \\
0.25 & 1.0000 & 0.9961 & 0.9990 \\
\midrule\multicolumn{4}{c}{$\rho_{12} = 0$} \\
0 & 0.0478 & 0.03818 & 0.01914 \\
0.01 & 0.3816 & 0.3355 & 0.2459 \\
0.05 & 0.9875 & 0.9819 & 0.9670 \\
0.1 & 0.9999 & 0.9997 & 0.9990 \\
0.25 & 1.0000 & 0.9999 & 0.9994 \\
\midrule\multicolumn{4}{c}{$\rho_{12} = 0.5$} \\
0 & 0.03165 & 0.05479 & 0.01202 \\
0.01 & 0.2879 & 0.3144 & 0.1702 \\
0.05 & 0.9655 & 0.9578 & 0.9191 \\
0.1 & 0.9988 & 0.9985 & 0.9967 \\
0.25 & 1.0000 & 1.0000 & 1.0000 \\
\midrule\multicolumn{4}{c}{$\rho_{12} = 0.75$} \\
0 & 0.01686 & 0.0588 & 0.006924 \\
0.01 & 0.1721 & 0.2313 & 0.08815 \\
0.05 & 0.8592 & 0.8464 & 0.7527 \\
0.1 & 0.9821 & 0.9784 & 0.9581 \\
0.25 & 0.9996 & 0.9994 & 0.9990 \\
\midrule\multicolumn{4}{c}{$\rho_{12} = 0.95$} \\
0 & 0.002107 & 0.06493 & 0.001706 \\
0.01 & 0.02321 & 0.1231 & 0.01021 \\
0.05 & 0.3389 & 0.3702 & 0.1745 \\
0.1 & 0.5843 & 0.5847 & 0.4011 \\
0.25 & 0.8056 & 0.8030 & 0.6670 \\
\midrule\bottomrule\end{tabular}
\end{table}%

\newpage%

\thispagestyle{empty}%
%

\begin{table}[ht]
\centering\caption{Results for the empirical illustration}
\begin{tabular}{lccccccc}
\hline& AORD & BFX & IBEX & MXX & NSEI & OMX \\
\hline$\delta$ & 0.03 & 0.05 & 0.08 & 0.25 & 0.35 & 0.27 \\
$\beta_{ARCH}$ & 0.01 & 0.01 & 0.00 & 0.08 & 0.16 & 0.00 \\
$\beta_{RV}$ & 0.09 & 0.63 & 0.62 & 0.30 & 0.39 & 0.49 \\
$\beta_{W}$ & 0.69 & 0.79 & 0.52 & 0.37 & 0.38 & 0.58 \\
$\beta_{M}$ & 0.34 & 0.00 & 0.17 & 0.72 & 0.22 & 0.05 \\
$\beta_{SPX}$ & 0.19 & 0.07 & 0.09 & 0.07 & 0.15 & 0.06 \\
\hline LR($\mathsf{H}_{ARCH}$) & 0.66 & 0.66 & 0.00 & 16.78 & 106.42 & 0.00 \\
Uniform CV & 6.58 & 6.66 & 8.39 & 16.98 & 20.44 & 18.50 \\
\hline LR($\mathsf{H}_{SPX}%
$) & 441.02 & 23.24 & 21.79 & 22.07 & 59.99 & 6.05 \\
Uniform CV & 7.57 & 7.29 & 8.52 & 16.95 & 20.21 & 18.76 \\
\hline Observations & 4757 & 4802 & 4772 & 4729 & 4670 & 3259 \\
\hline\end{tabular}
\label{tab:results_transposed}
\end{table}%

\appendix%

\setcounter{page}{1}%

\newpage

\setcounter{section}{0}

\section*{Supplemental Appendix}

\label{appendix}

\setcounter{equation}{1}

\renewcommand{\theequation}{A.\arabic{equation}}

\section{Proofs}

\subsection{Proofs of main results\label{sec: proofs}}

\subsubsection*{\emph{Proof of Lemma \ref{lem: LR}}}

We first introduce some notation. Let $b^{(1)}=(b_{1}^{(1)},\dots,b_{d_{\beta
}}^{(1)})^{\prime}$ and $b^{(2)}=(b_{1}^{(2)},\dots,b_{d_{\beta}}%
^{(2)})^{\prime}$ satisfy%
\[
b_{i}^{(1)}=\mathbb{I}(b_{i}<\infty)b_{i},\text{ }b_{i}^{(2)}=\mathbb{I(}%
b_{i}=\infty)b_{i}%
\]
for $i=1,\dots,d_{\beta}$, and where we work under the convention that
$0\times\infty=0$. Note that $b=b^{(1)}+b^{(2)}$. Likewise, let $\beta
_{n}^{(1)}=(\beta_{n,1}^{(1)},\dots,\beta_{n,d_{\beta}}^{(1)})^{\prime}$ and
$\beta_{n}^{(2)}=(\beta_{n,1}^{(2)},\dots,\beta_{n,d_{\beta}}^{(2)})^{\prime}$
satisfy%
\[
\beta_{n,i}^{(1)}=\mathbb{I(}b_{i}<\infty)\beta_{n,i},\quad\beta_{n,i}%
^{(2)}=\mathbb{I(}b_{i}=\infty)\beta_{n,i}%
\]
for $i=1,\dots,d_{\beta}$. Define%
\begin{equation}
\theta_{n}^{(1)}=(0_{d_{\gamma}}^{\prime},\beta_{n}^{(1)\prime},0_{d_{\delta}%
}^{\prime})^{\prime}\quad\text{and\quad}\theta_{n}^{(2)}=(\gamma_{0}^{\prime
},\beta_{n}^{(2)\prime},\delta_{0}^{\prime})^{\prime}.
\label{eq:def:theta_n_1_2}%
\end{equation}
In particular, $\theta_{n}=\theta_{n}^{(1)}+\theta_{n}^{(2)}$, and we have
that under any $\{\psi_{n}\}$ sequence,
\begin{equation}
\sqrt{n}\theta_{n}^{(1)}\rightarrow(0_{d_{\gamma}}^{\prime},b^{(1)}%
,0_{d_{\delta}}^{\prime})^{\prime}\equiv\tau\label{eq:limit:tau}%
\end{equation}
and%
\begin{equation}
\sqrt{n}(\Theta-\theta_{n}^{(2)})\rightarrow\tilde{\Lambda}\quad
\text{and\quad}\sqrt{n}(\Theta_{_{\mathsf{H}_{0}}}-\theta_{n}^{(2)}%
)\rightarrow\tilde{\Lambda}_{0} \label{eq:limit:Lambda_tilde}%
\end{equation}
where $\Theta_{_{\mathsf{H}_{0}}}\equiv\{\theta\in\Theta:\gamma=\gamma_{0}\}$
and $\tilde{\Lambda}$ and $\tilde{\Lambda}_{0}$ are convex cones (with zero
vertex) given respectively by%
\begin{equation}
\tilde{\Lambda}=\Lambda_{\gamma}\times\tilde{\Lambda}_{\beta}\times
\Lambda_{\delta}\quad\text{and\quad}\tilde{\Lambda}_{0}=\{0\}^{d_{\gamma}%
}\times\tilde{\Lambda}_{\beta}\times\Lambda_{\delta},
\label{eq:def:Lambda_tilde}%
\end{equation}
for $\Lambda_{\delta}=%
\mathbb{R}
^{d_{\delta}}$ and $\tilde{\Lambda}_{\beta}=\tilde{\Lambda}_{\beta,1}%
\times\cdots\times\tilde{\Lambda}_{\beta,d_{\beta}}$, with%
\[
\tilde{\Lambda}_{\beta,i}=\left\{
\begin{array}
[c]{lc}%
\mathbb{R}
_{+} & \text{if }b_{i}<\infty\\%
\mathbb{R}%
& \text{if }b_{i}=\infty
\end{array}
\right.  ,\quad i=1,\dots,d_{\beta}\text{.}%
\]
As in Andrews (1999), we note that
\[
L_{n}\left(  \theta\right)  =L_{n}\left(  \theta_{n}\right)  +\frac{1}{2}%
Z_{n}^{\prime}\Omega_{n}Z_{n}-\frac{1}{2}\Vert\sqrt{n}(\theta-\theta
_{n})-Z_{n}\Vert_{\Omega_{n}}^{2}+R_{n}\left(  \theta\right)  ,
\]
with%
\begin{align}
\Omega_{n}  &  \equiv-n^{-1}\frac{\partial^{2}L_{n}(\theta_{n})}%
{\partial\theta\partial\theta^{\prime}},\label{eq:def:Omega_n}\\
Z_{n}  &  =\Omega_{n}^{-1}n^{-1/2}\frac{\partial L_{n}(\theta_{n})}%
{\partial\theta},\label{eq:def:Z_n}\\
\Vert\lambda-Z_{n}\Vert_{\Omega_{n}}^{2}  &  =\left(  \lambda-Z_{n}\right)
^{\prime}\Omega_{n}\left(  \lambda-Z_{n}\right)  ,\quad\lambda\in%
\mathbb{R}
^{d_{\theta}}. \label{eq:def:quad_form}%
\end{align}
Hence,%
\begin{align*}
\mathsf{LR}_{n}  &  =2(L_{n}(\hat{\theta}_{n})-L_{n}(\tilde{\theta}_{n}))\\
&  =\Vert\sqrt{n}(\tilde{\theta}_{n}-\theta_{n})-Z_{n}\Vert_{\Omega_{n}}%
^{2}-\Vert\sqrt{n}(\hat{\theta}_{n}-\theta_{n})-Z_{n}\Vert_{\Omega_{n}}%
^{2}+2\left(  R_{n}(\hat{\theta}_{n})+R_{n}(\tilde{\theta}_{n})\right) \\
&  =\Vert\sqrt{n}(\tilde{\theta}_{n}-\theta_{n})-Z_{n}\Vert_{\Omega_{n}}%
^{2}-\Vert\sqrt{n}(\hat{\theta}_{n}-\theta_{n})-Z_{n}\Vert_{\Omega_{n}}%
^{2}+o_{p}\left(  1\right)  ,
\end{align*}
where the last equality follows by Assumptions \ref{ass:consistency}%
--\ref{ass:derivatives}. We seek to show that, jointly,%
\[
\Vert\sqrt{n}(\tilde{\theta}_{n}-\theta_{n})-Z_{n}\Vert_{\Omega_{n}}%
^{2}\overset{d}{\rightarrow}\inf_{\lambda\in\tilde{\Lambda}_{0}}\Vert
\lambda-(Z+\tau)\Vert_{\Omega_{0}}^{2}%
\]
and%
\[
\Vert\sqrt{n}(\hat{\theta}_{n}-\theta_{n})-Z_{n}\Vert_{\Omega_{n}}%
^{2}\overset{d}{\rightarrow}\inf_{\lambda\in\tilde{\Lambda}}\Vert
\lambda-(Z+\tau)\Vert_{\Omega_{0}}^{2}.
\]
Given this convergence, the limiting distribution, $\mathcal{L}_{\infty}%
(b,b)$, of $\mathsf{LR}_{n}$ is then derived by standard arguments, using the
structure of $\tilde{\Lambda}$ and $\tilde{\Lambda}_{0}$ and that
$\Lambda_{\delta}=%
\mathbb{R}
^{d_{\delta}}$:
\begin{align*}
&  \inf_{\lambda\in\tilde{\Lambda}_{0}}\Vert\lambda-(Z+\tau)\Vert_{\Omega_{0}%
}^{2}-\inf_{\lambda\in\tilde{\Lambda}}\Vert\lambda-(Z+\tau)\Vert_{\Omega_{0}%
}^{2}\\
&  =\inf_{\lambda\in\{0\}^{d_{\gamma}}\times\tilde{\Lambda}_{\beta}%
\times\Lambda_{\delta}}\Vert\lambda-(Z+\tau)\Vert_{\Omega_{0}}^{2}%
-\inf_{\lambda\in\Lambda_{\gamma}\times\tilde{\Lambda}_{\beta}\times
\Lambda_{\delta}}\Vert\lambda-(Z+\tau)\Vert_{\Omega_{0}}^{2}\\
&  =\inf_{\lambda\in\{0\}^{d_{\gamma}}\times\tilde{\Lambda}_{\beta}}%
\Vert\lambda-H(Z+\tau)\Vert_{(H\Omega_{0}^{-1}H^{\prime})^{-1}}^{2}%
-\inf_{\lambda\in\Lambda_{\gamma}\times\tilde{\Lambda}_{\beta}}\Vert
\lambda-H(Z+\tau)\Vert_{(H\Omega_{0}^{-1}H^{\prime})^{-1}}^{2}\\
&  =\inf_{\lambda\in\{0\}^{d_{\gamma}}\times\Lambda_{\beta}(b)}\Vert
\lambda-HZ\Vert_{(H\Omega_{0}^{-1}H^{\prime})^{-1}}^{2}-\inf_{\lambda
\in\Lambda_{\gamma}\times\Lambda_{\beta}(b)}\Vert\lambda-HZ\Vert_{(H\Omega
_{0}^{-1}H^{\prime})^{-1}}^{2}\\
&  =\mathcal{L}_{\infty}(b,b).
\end{align*}
Let us focus on the convergence of $\Vert\sqrt{n}(\hat{\theta}_{n}-\theta
_{n})-Z_{n}\Vert_{\Omega_{n}}^{2}$, noting that the convergence of $\Vert
\sqrt{n}(\tilde{\theta}_{n}-\theta_{n})-Z_{n}\Vert_{\Omega_{n}}^{2}$ follows
by similar arguments, and that the convergence holds jointly, as $\Vert
\sqrt{n}(\hat{\theta}_{n}-\theta_{n})-Z_{n}\Vert_{\Omega_{n}}^{2}$ and
$\Vert\sqrt{n}(\tilde{\theta}_{n}-\theta_{n})-Z_{n}\Vert_{\Omega_{n}}^{2}$ are
functions of the same data $\{W_{t}\}$. It suffices to show the following
three properties:

\begin{enumerate}
\item With $\hat{\theta}_{q,n}$ satisfying $\Vert\sqrt{n}(\hat{\theta}%
_{q,n}-\theta_{n})-Z_{n}\Vert_{\Omega_{n}}^{2}=\inf_{\theta\in\Theta}%
\Vert\sqrt{n}(\theta-\theta_{n})-Z_{n}\Vert_{\Omega_{n}}^{2}$, it holds that
\[
\Vert\sqrt{n}(\hat{\theta}_{n}-\theta_{n})-Z_{n}\Vert_{\Omega_{n}}^{2}%
=\Vert\sqrt{n}(\hat{\theta}_{q,n}-\theta_{n})-Z_{n}\Vert_{\Omega_{n}}%
^{2}+o_{p}(1).
\]

\item It holds that
\[
\Vert\sqrt{n}(\hat{\theta}_{q,n}-\theta_{n})-Z_{n}\Vert_{\Omega_{n}}^{2}%
=\inf_{\theta\in\Theta}\Vert\sqrt{n}(\theta-\theta_{n})-Z_{n}\Vert_{\Omega
_{n}}^{2}=\inf_{\lambda\in\tilde{\Lambda}}\Vert\lambda-\sqrt{n}\theta
_{n}^{(1)}-Z_{n}\Vert_{\Omega_{n}}^{2}+o_{p}(1).
\]

\item It holds that
\[
\inf_{\lambda\in\tilde{\Lambda}}\Vert\lambda-\sqrt{n}\theta_{n}^{(1)}%
-Z_{n}\Vert_{\Omega_{n}}^{2}\overset{d}{\rightarrow}\inf_{\lambda\in
\tilde{\Lambda}}\Vert\lambda-(Z+\tau)\Vert_{\Omega_{0}}^{2}.
\]

\end{enumerate}

These properties follow from Lemmas \ref{lem:min_of_quadform}%
--\ref{lem:limit_of_min_cone}.\hfill$\square$

\subsubsection*{\emph{Proof of Lemma \ref{lem:monotonicity}}}

With $\underline{b}\leq b\leq\bar{b},$ it holds that $\Lambda_{\beta
}(\underline{b})\subset\Lambda_{\beta}(b)\subset\Lambda_{\beta}(\bar{b})$.
This implies that
\[
\inf_{\lambda\in\{0\}^{d_{\gamma}}\times\Lambda_{\beta}(\underline{b}%
)}Q\left(  \lambda\right)  \geq\inf_{\lambda\in\{0\}^{d_{\gamma}}\times
\Lambda_{\beta}(b)}Q\left(  \lambda\right)  ,
\]
and
\[
\inf_{\lambda\in\Lambda_{\gamma}\times\Lambda_{\beta}(b)}Q\left(
\lambda\right)  \geq\inf_{\lambda\in\Lambda_{\gamma}\times\Lambda_{\beta}%
(\bar{b})}Q\left(  \lambda\right)  .
\]
Hence,%
\begin{align*}
\mathcal{L}_{\infty}(b,b)  &  =\inf_{\lambda\in\{0\}^{d_{\gamma}}\times
\Lambda_{\beta}(b)}Q\left(  \lambda\right)  -\inf_{\lambda\in\Lambda_{\gamma
}\times\Lambda_{\beta}(b)}Q\left(  \lambda\right)  \leq\inf_{\lambda
\in\{0\}^{d_{\gamma}}\times\Lambda_{\beta}(\underline{b})}Q\left(
\lambda\right)  -\inf_{\lambda\in\Lambda_{\gamma}\times\Lambda_{\beta}(\bar
{b})}Q\left(  \lambda\right) \\
&  =\mathcal{L}_{\infty}(\underline{b},\bar{b}),
\end{align*}
as required.\hfill$\square$

\subsubsection*{\emph{Proof of Lemma \ref{lem:Ketz}}}

By a Taylor-type expansion at $\theta_{n}$,%
\begin{align*}
\sqrt{n}\left(  \check{\theta}_{n}-\theta_{n}\right)   &  =\left[
I_{d_{\theta}}-\left(  \frac{\partial^{2}L_{n}(\hat{\theta}_{n})}%
{\partial\theta\partial\theta^{\prime}}\right)  ^{-1}\frac{\partial^{2}%
L_{n}(\theta_{n})}{\partial\theta\partial\theta^{\prime}}\right]  \sqrt
{n}(\hat{\theta}_{n}-\theta_{n})\\
&  -\left(  n^{-1}\frac{\partial^{2}L_{n}(\hat{\theta}_{n})}{\partial
\theta\partial\theta^{\prime}}\right)  ^{-1}n^{-1/2}\frac{\partial
L_{n}(\theta_{n})}{\partial\theta}+o_{p}\left(  1\right)
\overset{d}{\rightarrow}N(0,\Omega_{0}^{-1}\Sigma_{0}\Omega_{0}^{-1})
\end{align*}
where we have used Assumptions \ref{ass:consistency}--\ref{ass:score} and
Remark \ref{rem:convergence_rate}. This proves (\ref{eq:AN_beta_bar}). In
addition, (\ref{eq:cons_Sigma_beta}) follows directly from Assumptions
\ref{ass:consistency}--\ref{ass:hessian} and \ref{ass:covariance-consistency}%
.\hfill$\square$

\subsubsection*{\emph{Proof of Theorem \ref{thm:asysize control}}}

For the parameter space $\Psi$, standard subsequencing arguments
(e.g.,~Andrews and Guggenberger, 2009, and McCloskey, 2017) provide that
showing
\begin{equation}
\lim_{n\rightarrow\infty}\mathbb{P}_{\psi_{n}}\left(  \mathsf{LR}_{n}%
\geq\mathsf{CV}_{1-\alpha+\eta,n}(b_{L,n},b_{U,n})\right)  \leq\alpha
\label{eq:unif_0}%
\end{equation}
under all $\{\psi_{n}\}$ sequences in $\Psi$ satisfying $\psi_{n}%
\rightarrow\psi_{0}$ and (\ref{eq:def_drifting_sequence}) is sufficient for
proving the statement of the theorem. Consider any such sequence; then, we
have
\begin{align}
&  \mathbb{P}_{\psi_{n}}\left(  \mathsf{LR}_{n}\geq\mathsf{CV}_{1-\alpha
+\eta,n}(b_{L,n},b_{U,n})\right) \nonumber\\
&  =\mathbb{P}_{\psi_{n}}\left(  \mathsf{LR}_{n}\geq\mathsf{CV}_{1-\alpha
+\eta,n}(b_{L,n},b_{U,n})\geq\mathsf{CV}_{1-\alpha+\eta,n}(\sqrt{n}\beta
_{n},\sqrt{n}\beta_{n})\right) \nonumber\\
&  \quad+\mathbb{P}_{\psi_{n}}\left(  \mathsf{LR}_{n}\geq\mathsf{CV}%
_{1-\alpha+\eta,n}(\sqrt{n}\beta_{n},\sqrt{n}\beta_{n})>\mathsf{CV}%
_{1-\alpha+\eta,n}(b_{L,n},b_{U,n})\right) \nonumber\\
&  \quad+\mathbb{P}_{\psi_{n}}\left(  \mathsf{CV}_{1-\alpha+\eta,n}(\sqrt
{n}\beta_{n},\sqrt{n}\beta_{n})>\mathsf{LR}_{n}\geq\mathsf{CV}_{1-\alpha
+\eta,n}(b_{L,n},b_{U,n})\right) \nonumber\\
&  \leq\mathbb{P}_{\psi_{n}}\left(  \mathsf{LR}_{n}\geq\mathsf{CV}%
_{1-\alpha+\eta,n}(\sqrt{n}\beta_{n},\sqrt{n}\beta_{n})\right)
\label{eq:unif_1}\\
&  \quad+\mathbb{P}_{\psi_{n}}\left(  \mathsf{CV}_{1-\alpha+\eta,n}(\sqrt
{n}\beta_{n},\sqrt{n}\beta_{n})>\mathsf{CV}_{1-\alpha+\eta,n}(b_{L,n}%
,b_{U,n})\right)  . \nonumber\label{eq:unif_2}%
\end{align}

\noindent Note the following:

\begin{enumerate}
\item[(a)] the distribution function of $\mathcal{L}_{\infty,n}(\sqrt{n}%
\beta_{n},\sqrt{n}\beta_{n})$ converges in probability to the distribution
function of $\mathcal{L}_{\infty}(b,b)$ by Assumptions \ref{ass:consistency}%
--\ref{ass:hessian} and \ref{ass:covariance-consistency} and the continuous
mapping theorem;

\item[(b)] $\mathsf{LR}_{n}\overset{d}{\rightarrow}\mathcal{L}_{\infty}(b,b)$
by Lemma \ref{lem: LR};

\item[(c)] for any $b\in\lbrack0,\infty]^{d_{\beta}}$, $\mathcal{L}_{\infty
}(b,b)$ is an absolutely continuous random variable with support $[0,\infty)$.
\end{enumerate}

\noindent Therefore, Lemma 5(ii) of Andrews and Guggenberger (2010) implies
\begin{equation}
\mathbb{P}_{\psi_{n}}\left(  \mathsf{LR}_{n}\geq\mathsf{CV}_{1-\alpha+\eta
,n}(\sqrt{n}\beta_{n},\sqrt{n}\beta_{n})\right)  \rightarrow\mathbb{P}\left(
\mathcal{L}_{\infty}(b,b)\geq\mathsf{CV}_{1-\alpha+\eta}(b,b)\right)
=\alpha-\eta, \label{eq:unif_3}%
\end{equation}
where $\mathsf{CV}_{1-\alpha+\eta}(b,b)$ denotes the $1-\alpha+\eta$ quantile
of $\mathcal{L}_{\infty}(b,b)$. For $q_{1-\eta}$ equal to the $\left(
1-\eta\right)  $-quantile of $\max_{i=1,\dots,d_{\beta}}|Z_{\beta,i}|$ with
$Z_{\beta}\overset{d}{=}N(0,\Omega_{\beta})$ and $\Omega_{\beta}%
=\operatorname*{diag}\left(  \Sigma_{\beta}\right)  ^{-1/2}\Sigma_{\beta
}\operatorname*{diag}\left(  \Sigma_{\beta}\right)  ^{-1/2}$,
\begin{align}
&  \mathbb{P}_{\psi_{n}}\left(  \mathsf{CV}_{1-\alpha+\eta,n}(\sqrt{n}%
\beta_{n},\sqrt{n}\beta_{n})>\mathsf{CV}_{1-\alpha+\eta,n}(b_{L,n}%
,b_{U,n})\right) \nonumber\\
&  =1-\mathbb{P}_{\psi_{n}}\left(  \mathsf{CV}_{1-\alpha+\eta,n}(\sqrt{n}%
\beta_{n},\sqrt{n}\beta_{n})\leq\mathsf{CV}_{1-\alpha+\eta,n}(b_{L,n}%
,b_{U,n})\right) \nonumber\\
&  \leq1-\mathbb{P}_{\psi_{n}}\left(  b_{L,n}\leq\sqrt{n}\beta_{n}\leq
b_{U,n}\right) \nonumber\\
&  =1-\mathbb{P}_{\psi_{n}}\left(  \bar{b}_{L,n}\leq\sqrt{n}\beta_{n}\leq
b_{U,n}\right) \nonumber\\
&  \leq1-\mathbb{P}_{\psi_{n}}\left(  \bar{b}_{L,n}\leq\sqrt{n}\beta_{n}%
\leq\bar{b}_{U,n}\right) \nonumber\\
&  =1-\mathbb{P}_{\psi_{n}}\left(  -\hat{q}_{1-\eta,n}\operatorname*{diagv}%
(\hat{\Sigma}_{\beta,n})^{1/2}\leq\sqrt{n}(\bar{\beta}_{n}-\beta_{n})\leq
\hat{q}_{1-\eta,n}\operatorname*{diagv}(\hat{\Sigma}_{\beta,n})^{1/2})\right)
\nonumber\\
&  =1-\mathbb{P}_{\psi_{n}}\left(  -\hat{q}_{1-\eta,n}\leq\frac{\sqrt{n}%
(\bar{\beta}_{n,i}-\beta_{n,i})}{\sqrt{\hat{\Sigma}_{\beta,n,ii}}}\leq\hat
{q}_{1-\eta,n}\text{ for all }i=1,\ldots,d_{\beta}\right) \nonumber\\
&  =1-\mathbb{P}_{\psi_{n}}\left(  \max_{i=1,\ldots,d_{\beta}}\left\vert
\frac{\sqrt{n}(\bar{\beta}_{n,i}-\beta_{n,i})}{\sqrt{\hat{\Sigma}_{\beta
,n,ii}}}\right\vert \leq\hat{q}_{1-\eta,n}\right) \nonumber\\
&  \rightarrow1-\mathbb{P}\left(  \max_{i=1,\ldots,d_{\beta}}\left\vert
Z_{\beta,i}\right\vert \leq q_{1-\eta}\right)  =\eta, \label{eq:unif_4}%
\end{align}
where the inequalities inside of probabilities are evaluated element-wise
across vectors, the first inequality follows from Lemma \ref{lem:monotonicity}%
, the second equality follows from $\beta_{n}\geq0$ and the convergence
follows from Assumption \ref{ass:AN_estimator} and Lemma 5(ii) of Andrews and
Guggenberger (2010).

Together, (\ref{eq:unif_1})--(\ref{eq:unif_4}) imply (\ref{eq:unif_0}), and
therefore the statement of the theorem. $\square$

\subsection{Technical lemmas for proving main results}

\begin{lemma}
\label{lem:min_of_quadform}Let $\hat{\theta}_{q,n}$ satisfy $\Vert\sqrt
{n}(\hat{\theta}_{q,n}-\theta_{n})-Z_{n}\Vert_{\Omega_{n}}^{2}=\inf_{\theta
\in\Theta}\Vert\sqrt{n}(\theta-\theta_{n})-Z_{n}\Vert_{\Omega_{n}}^{2},$ with
$\Omega_{n}$ given by (\ref{eq:def:Omega_n}) and $Z_{n}$ given by
(\ref{eq:def:Z_n}). Under Assumptions \ref{ass:consistency}-\ref{ass:score}
and any sequence $\left\{  \psi_{n}\right\}  $ satisfying $\psi_{n}%
\rightarrow\psi_{0}$ and (\ref{eq:def_drifting_sequence}),%
\[
\Vert\sqrt{n}(\hat{\theta}_{n}-\theta_{n})-Z_{n}\Vert_{\Omega_{n}}^{2}%
=\Vert\sqrt{n}(\hat{\theta}_{q,n}-\theta_{n})-Z_{n}\Vert_{\Omega_{n}}%
^{2}+o_{p}(1).
\]

\end{lemma}

\noindent\textsc{Proof}. The result follows directly from arguments given in
Andrews (1999, proof of Theorem 2) with $\theta_{0}$ replaced by $\theta_{n}%
$.\hfill$\square$

\begin{lemma}
\label{lem:min_cone}Under Assumptions \ref{ass:hessian}-\ref{ass:score} and
any sequence $\left\{  \psi_{n}\right\}  $ satisfying $\psi_{n}\rightarrow
\psi_{0}$ and (\ref{eq:def_drifting_sequence}), it holds that
\[
\inf_{\theta\in\Theta}\Vert\sqrt{n}(\theta-\theta_{n})-Z_{n}\Vert_{\Omega_{n}%
}^{2}=\inf_{\lambda\in\tilde{\Lambda}}\Vert\lambda-\sqrt{n}\theta_{n}%
^{(1)}-Z_{n}\Vert_{\Omega_{n}}^{2}+o_{p}(1),
\]
where $\Omega_{n}$ given by (\ref{eq:def:Omega_n}), $Z_{n}$ given by
(\ref{eq:def:Z_n}), $\tilde{\Lambda}$ is given by (\ref{eq:def:Lambda_tilde})
and $\theta_{n}^{(1)}$ is given by (\ref{eq:def:theta_n_1_2}).
\end{lemma}

\noindent\textsc{Proof}. By definition, using (\ref{eq:def:theta_n_1_2}),
\[
\Vert\sqrt{n}(\theta-\theta_{n})-Z_{n}\Vert_{\Omega_{n}}^{2}=\Vert\sqrt
{n}(\theta-\theta_{n}^{(2)})-(Z_{n}+\sqrt{n}\theta_{n}^{(1)})\Vert_{\Omega
_{n}}^{2}.
\]
Hence,%
\begin{align*}
\inf_{\theta\in\Theta}\Vert\sqrt{n}(\theta-\theta_{n})-Z_{n}\Vert_{\Omega_{n}%
}^{2}  &  =\inf_{\theta\in\Theta}\Vert\sqrt{n}(\theta-\theta_{n}^{(2)}%
)-(Z_{n}+\sqrt{n}\theta_{n}^{(1)})\Vert_{\Omega_{n}}^{2}\\
&  =\inf_{\lambda\in\sqrt{n}(\Theta-\theta_{n}^{(2)})}\Vert\lambda
-(Z_{n}+\sqrt{n}\theta_{n}^{(1)})\Vert_{\Omega_{n}}^{2}.
\end{align*}
By Assumptions \ref{ass:hessian}-\ref{ass:score} and (\ref{eq:limit:tau}), we
have that $Z_{n}+\sqrt{n}\theta_{n}^{(1)}=O_{p}(1)$. Hence, the result now
follows using the fact that $\sqrt{n}(\Theta-\theta_{n}^{(2)})$ contains zero
for all $n\geq1$ and Silvapulle and Sen (2005, Corollary 4.7.5.1 and the
comments on p.~194).\hfill$\square$

\begin{lemma}
\label{lem:limit_of_min_cone}Under Assumptions \ref{ass:hessian}%
-\ref{ass:score} and any sequence $\left\{  \psi_{n}\right\}  $ satisfying
$\psi_{n}\rightarrow\psi_{0}$ and (\ref{eq:def_drifting_sequence}), it holds
that%
\[
\inf_{\lambda\in\tilde{\Lambda}}\Vert\lambda-(Z_{n}+\sqrt{n}\theta_{n}%
^{(1)})\Vert_{\Omega_{n}}^{2}\overset{d}{\rightarrow}\inf_{\lambda\in
\tilde{\Lambda}}\Vert\lambda-(Z+\tau)\Vert_{\Omega_{0}}^{2},
\]
where $\Omega_{n}$ given by (\ref{eq:def:Omega_n}), $Z_{n}$ given by
(\ref{eq:def:Z_n}), $\tilde{\Lambda}$ is given by (\ref{eq:def:Lambda_tilde}),
$\theta_{n}^{(1)}$ is given by (\ref{eq:def:theta_n_1_2}), and $\tau$ is
defined in (\ref{eq:limit:tau}).
\end{lemma}

\noindent\textsc{Proof}. Since $(Z_{n}+\sqrt{n}\theta_{n}^{(1)})=O_{p}(1)$ and
$\Omega_{n}=\Omega_{0}+o_{p}(1)$ by Assumptions \ref{ass:hessian}%
-\ref{ass:score}, we have by Silvapulle and Sen (2005, Lemma 4.10.2.2) that
\[
\inf_{\lambda\in\tilde{\Lambda}}\Vert\lambda-(Z_{n}+\sqrt{n}\theta_{n}%
^{(1)})\Vert_{\Omega_{n}}^{2}=\inf_{\lambda\in\tilde{\Lambda}}\Vert
\lambda-(Z_{n}+\sqrt{n}\theta_{n}^{(1)})\Vert_{\Omega_{0}}^{2}+o_{p}(1).
\]
By Silvapulle and Sen (2005, Corollary 4.7.5.2) and the fact that $Z_{n}%
+\sqrt{n}\theta_{n}^{(1)}\overset{d}{\rightarrow}Z+\tau$, we have that
$\inf_{\lambda\in\tilde{\Lambda}}\Vert\lambda-(Z_{n}+\sqrt{n}\theta_{n}%
^{(1)})\Vert_{\Omega_{0}}^{2}\overset{d}{\rightarrow}\inf_{\lambda\in
\tilde{\Lambda}}\Vert\lambda-(Z+\tau)\Vert_{\Omega_{0}}^{2}.$\hfill$\square$

\subsection{Proofs and lemmas related to linear regression example}

\subsubsection*{\emph{Proof of Proposition \ref{prop:reg ass verification}}}

Starting with Assumption \ref{ass:consistency}, note that Assumptions LinIID
\ref{ass:lin-iid}.1--3 and a weak LLN for row-wise i.i.d. random variables
imply that $\Vert S_{xx}-\mathbb{E}_{\psi_{n}}[x_{t}x_{t}^{\prime}%
]\Vert\overset{p}{\rightarrow}0$ and $\Vert S_{x\varepsilon}-\mathbb{E}%
_{\psi_{n}}[x_{t}\varepsilon_{t}]\Vert\overset{p}{\rightarrow}0$ (all
convergence statements that follow are thus understood to be under any
sequence $\left\{  \psi_{n}\right\}  $ satisfying $\psi_{n}\rightarrow\psi
_{0}$ and (\ref{eq:def_drifting_sequence})). Since $x_{t}$ under $\psi_{n}$
converges in distribution to $x_{t}$ under $\psi_{0}$ and $\max_{j}E_{\psi
}[|x_{t,j}|^{2+\nu}]\leq c$ for all $\psi\in\Psi$, we have that $\mathbb{E}%
_{\psi_{n}}[x_{t}x_{t}^{\prime}]\rightarrow\mathbb{E}_{\psi_{0}}[x_{t}%
x_{t}^{\prime}]=\Omega_{0},$ such that $S_{xx}\overset{p}{\rightarrow}%
\Omega_{0}$. Likewise, $\mathbb{E}_{\psi}[x_{t}\varepsilon_{t}]=0$ for all
$\psi\in\Psi,$ such that $S_{x\varepsilon}\overset{p}{\rightarrow}0$. Clearly,
these convergence properties imply that $S_{xx}$ is invertible with
probability approaching one, such that%
\begin{equation}
\hat{\theta}_{LS}-\theta_{n}=S_{xx}^{-1}S_{x\varepsilon}=o_{p}(1),
\label{eq:LS consistency}%
\end{equation}
and%
\[
\Vert\hat{\theta}_{LS}-\theta_{n}\Vert_{S_{xx}}^{2}=(\hat{\theta}_{LS}%
-\theta_{n})^{\prime}S_{xx}(\hat{\theta}_{LS}-\theta_{n})=o_{p}(1).
\]
We then have by the triangle inequality%
\[
\Vert\hat{\theta}_{n}-\theta_{n}\Vert_{S_{xx}}\leq\Vert\hat{\theta}_{n}%
-\hat{\theta}_{LS}\Vert_{S_{xx}}+\Vert\hat{\theta}_{LS}-\theta_{n}%
\Vert_{S_{xx}}\leq2\Vert\hat{\theta}_{LS}-\theta_{n}\Vert_{S_{xx}},
\]
where the second equality follows by noting that $\theta_{n}\in\Theta$ and
$\Vert\hat{\theta}_{n}-\hat{\theta}_{LS}\Vert_{S_{xx}}=\min_{\theta\in\Theta
}\Vert\theta-\hat{\theta}_{LS}\Vert_{S_{xx}}.$ We conclude that $\Vert
\hat{\theta}_{n}-\theta_{n}\Vert_{S_{xx}}=o_{p}(1),$ and hence that
$\hat{\theta}_{n}-\theta_{n}=o_{p}(1)$. By similar arguments we have that
$\tilde{\theta}_{n}-\theta_{n}=o_{p}(1)$.\newline Moving now to Assumption
\ref{ass:derivatives},~1.~clearly holds by the definition of $L_{n}(\cdot)$
and 2.~holds trivially since $-n^{-1}\partial^{2}L_{n}(\theta)/\partial
\theta\partial\theta^{\prime}=S_{xx}$ in this example. Assumption
\ref{ass:hessian} also holds by the fact that $-n^{-1}\partial^{2}L_{n}%
(\theta)/\partial\theta\partial\theta^{\prime}=S_{xx}$ and that $S_{xx}%
\overset{p}{\rightarrow}\Omega_{0}$. Finally, Assumption \ref{ass:score} holds
by Assumptions LinIID \ref{ass:lin-iid}.1--3, the Cram\'{e}r Wold device and a
Liapounov central limit theorem (CLT) for row-wise i.i.d. random variables
since $n^{-1/2}\partial L_{n}(\theta_{n})/\partial\theta=n^{1/2}%
S_{x\varepsilon}$ in this example.\hfill$\square$

\subsubsection*{\emph{Proof of Proposition \ref{prop:reg ass verification 2}}}

The proof that Assumption \ref{ass:covariance-consistency} holds is nearly
identical to those of White (1980, proof of Theorem 1). For Assumption
\ref{ass:AN_estimator}, note that under any sequence $\left\{  \psi
_{n}\right\}  $ satisfying $\psi_{n}\rightarrow\psi_{0}$ and
(\ref{eq:def_drifting_sequence}),
\[
\sqrt{n}(\hat{\theta}_{LS}-\theta_{n})=S_{xx}^{-1}\sqrt{n}S_{x\varepsilon
}\overset{d}{\rightarrow}N(0,\Omega_{0}^{-1}\Sigma_{0}\Omega_{0}^{-1})
\]
by the continuous mapping theorem and the fact that Assumptions
\ref{ass:hessian}--\ref{ass:score} hold. Furthermore, given that Assumptions
\ref{ass:hessian} and \ref{ass:covariance-consistency} hold, the continuous
mapping theorem implies $\hat{\Sigma}_{\beta,n}\overset{p}{\rightarrow}%
\Sigma_{\beta}$.\hfill\ $\square$

\begin{lemma}
\label{lem:nnls}Suppose that $S_{xx}$ is invertible. For any set
$\Lambda\subset\mathbb{R}^{1+d_{\beta}}$,
\[
\arg\inf_{\theta\in\Lambda}\sum_{t=1}^{n}\left(  y_{t}-x_{t}^{\prime}%
\theta\right)  ^{2}=\arg\inf_{\theta\in\Lambda}(\theta-\hat{\theta}%
_{LS})^{\prime}S_{xx}(\theta-\hat{\theta}_{LS}).
\]

\end{lemma}

\noindent\textsc{Proof.} After noting that for any $\theta$,%
\begin{align*}
\sum_{t=1}^{n}\left(  y_{t}-x_{t}^{\prime}\theta\right)  ^{2}  &  =\sum
_{t=1}^{n}(y_{t}-x_{t}^{\prime}\hat{\theta}_{LS}-x_{t}^{\prime}(\theta
-\hat{\theta}_{LS}))^{2}\\
&  =\sum_{t=1}^{n}(y_{t}-x_{t}^{\prime}\hat{\theta}_{LS})^{2}+\sum_{t=1}%
^{n}(x_{t}^{\prime}(\theta-\hat{\theta}_{LS}))^{2}\\
&  -2(\theta-\hat{\theta}_{LS})^{\prime}\underset{=0}{\underbrace{\sum
_{t=1}^{n}x_{t}(y_{t}-x_{t}^{\prime}\hat{\theta}_{LS})}}\\
&  =\sum_{t=1}^{n}(y_{t}-x_{t}^{\prime}\hat{\theta}_{LS})^{2}+n(\theta
-\hat{\theta}_{LS})^{\prime}S_{xx}(\theta-\hat{\theta}_{LS}),
\end{align*}
the result follows immediately.\hfill\ $\square$

\subsection{Proofs related to the ARCH example}

Below we prove that Assumptions \ref{ass:consistency}-\ref{ass:AN_estimator}
hold under Assumptions\ ARCH \ref{ass:ARCH-HL-stat-mixing}%
-\ref{ass:ARCH-HL-criterion}. Throughout, we make use of
\[
W_{t}=(y_{t}^{2},F_{t-1}^{\prime})^{\prime}\in\mathcal{W=}%
\mathbb{R}
_{+}\times\{1\}\times%
\mathbb{R}
_{+}^{d_{\beta}+d_{\gamma}}.
\]

\subsubsection*{\emph{Proof that Assumption \ref{ass:consistency} holds}}

Under under any sequence $\left\{  \psi_{n}\right\}  $ satisfying $\psi
_{n}\rightarrow\psi_{0}$ and (\ref{eq:def_drifting_sequence}), and using the
compactness of $\Theta$, to prove the convergence of $\hat{\theta}_{n}$, it
suffices to show that
\begin{equation}
\sup_{\theta\in\Theta}|n^{-1}L_{n}(\theta)-\mathcal{L}(\theta
)|\overset{p}{\rightarrow}0, \label{eq:criterion_ULLN}%
\end{equation}
with%
\begin{equation}
\mathcal{L}(\theta)=-\frac{1}{2}\mathbb{E}_{\psi_{0}}\left[  \log\sigma
_{t}^{2}(\theta)+\frac{y_{t}^{2}}{\sigma_{t}^{2}(\theta)}\right]
\label{eq:def_limiting_criterion}%
\end{equation}
and%
\begin{equation}
\mathcal{L}(\theta)\leq\mathcal{L}(\theta_{0})\text{ for any }\theta\in
\Theta\text{ with equality if and only if }\theta=\theta_{0}.
\label{eq:criterion-identification}%
\end{equation}
We start out by showing that (\ref{eq:criterion_ULLN}) by applying Lemma 11.3
of Andrews and Cheng (2013b)\footnote{A careful inspection of the proof of
that lemma shows that only strong mixing conditions as the ones stated in
Assumption ARCH \ref{ass:ARCH-HL-stat-mixing} are needed. In particular, the
proof makes use of a weak LLN for triangular arrays of strongly mixing
processes, which does not impose any rate of decay on the the mixing
coefficients.}. Recall that
\[
l_{t}(\theta)=-\frac{1}{2}\left(  \log\sigma_{t}^{2}(\theta)+\frac{y_{t}^{2}%
}{\sigma_{t}^{2}(\theta)}\right)  =-\frac{1}{2}\left(  \log g(F_{t-1}%
,\theta)+\frac{W_{t,1}}{g(F_{t-1},\theta)}\right)  ,
\]
with $g(F_{t-1},\theta):=F_{t-1}^{\prime}\theta$ and $W_{t,1}=y_{t}^{2}$, the
first entry of $W_{t}$. For any $w\in\mathcal{W}$, let $w_{1}$ denote the
first entry of $w$ and $w_{2}$ the column vector of the remaining entries,
that is, $w=(w_{1},w_{2}^{\prime})^{\prime}.$ Let
\[
s(w,\theta)=\log g(w,\theta)+\frac{w_{1}}{g(w_{2},\theta)}=\log\left(
w_{2}^{\prime}\theta\right)  +\frac{w_{1}}{w_{2}^{\prime}\theta}.
\]
For any $\theta_{1},\theta_{2}\in\Theta$, a mean value expansion gives%
\[
\log\left(  w_{2}^{\prime}\theta_{1}\right)  =\log\left(  w_{2}^{\prime}%
\theta_{2}\right)  +\frac{1}{w_{2}^{\prime}\theta^{\ast}}w_{2}^{\prime}%
(\theta_{1}-\theta_{2}),
\]
with $\theta^{\ast}\in\Theta$ between $\theta_{1}$ and $\theta_{2}$. Likewise,%
\[
\frac{w_{1}}{w_{2}^{\prime}\theta_{1}}=\frac{w_{1}}{w_{2}^{\prime}\theta_{2}%
}-\frac{w_{1}}{(w_{2}^{\prime}\theta^{\ast\ast})^{2}}w_{2}^{\prime}(\theta
_{1}-\theta_{2}),
\]
with $\theta^{\ast\ast}\in\Theta$ between $\theta_{1}$ and $\theta_{2}$. It
holds that $w_{2}^{\prime}\theta\geq\delta_{L}$ uniformly on $\mathcal{W}%
\times\Theta.$ Consequently, for all $\theta_{1},\theta_{2}\in\Theta$
\begin{align*}
|s(w,\theta_{1})-s(w,\theta_{2})|  &  =\left\vert \frac{1}{w_{2}^{\prime
}\theta^{\ast}}w_{2}^{\prime}(\theta_{1}-\theta_{2})-\frac{w_{1}}%
{(w_{2}^{\prime}\theta^{\ast\ast})^{2}}w_{2}^{\prime}(\theta_{1}-\theta
_{2})\right\vert \\
&  \leq\left(  \delta_{L}^{-1}+\delta_{L}^{-2}w_{1}\right)  \left\Vert
w_{2}\right\Vert \left\Vert \theta_{1}-\theta_{2}\right\Vert .
\end{align*}
With $M_{1}(w):=\left(  \delta_{L}^{-1}+\delta_{L}^{-2}w_{1}\right)
\left\Vert w_{2}\right\Vert $, we conclude that for any $\eta>0$%
\begin{equation}
|s(w,\theta_{1})-s(w,\theta_{2})|\leq M_{1}(w)\eta\label{eq:bound_M}%
\end{equation}
for all $\theta_{1},\theta_{2}\in\Theta$ and $w\in\mathcal{W}$ with
$\left\Vert \theta_{1}-\theta_{2}\right\Vert <\eta$. By Assumption ARCH
\ref{ass:ARCH-HL-criterion}, it holds that%
\begin{align*}
\mathbb{E}_{\psi}[M_{1}(W_{t})]  &  =\mathbb{E}_{\psi}\left[  \left(
\delta_{L}^{-1}+\delta_{L}^{-2}y_{t}^{2}\right)  \left\Vert F_{t-1}\right\Vert
\right] \\
&  \leq\delta_{L}^{-1}\mathbb{E}_{\psi}\left[  \left\Vert F_{t-1}\right\Vert
\right]  +\delta_{L}^{-2}\mathbb{E}_{\psi}\left[  y_{t}^{2}\left\Vert
F_{t-1}\right\Vert \right]  \leq\tilde{c}%
\end{align*}
for some constant $\tilde{c}\in(0,\infty)$ for all $\psi\in\Psi$. Moreover, we
have that for $\theta\in\Theta$%
\[
|s(W_{t},\theta)|\leq|\log(\delta_{L})|+\Vert F_{t-1}\Vert d_{\theta}%
(\delta_{U}+\beta_{U}+\gamma_{U})+\frac{y_{t}^{2}}{\delta_{L}},
\]
so using Assumption ARCH \ref{ass:ARCH-HL-criterion} again, we have that
\[
\mathbb{E}_{\psi}[\sup_{\theta\in\Theta}|s(W_{t},\theta)|^{1+\nu}]\leq
\tilde{c}%
\]
for some constants $\nu,\tilde{c}\in(0,\infty)$ for all $\psi\in\Psi$. We
conclude that
\begin{equation}
\mathbb{E}_{\psi}[\sup_{\theta\in\Theta}|s(W_{t},\theta)|^{1+\nu}%
]+\mathbb{E}_{\psi}[M_{1}(W_{t})]\leq\bar{C}, \label{eq:bound_likelihood_cont}%
\end{equation}
for some constants $\nu,\tilde{c}\in(0,\infty)$ for all $\psi\in\Psi$. Using
the fact that $l_{t}(\theta)=-s(W_{t},\theta)/2$ together with
(\ref{eq:bound_M}) and (\ref{eq:bound_likelihood_cont}), we have that
(\ref{eq:criterion_ULLN}) holds by Lemma 11.3 of Andrews and Cheng (2013b).
Condition (\ref{eq:criterion-identification}) holds by standard arguments and
Assumption \ref{ass:identification}. The properties (\ref{eq:criterion_ULLN}%
)-(\ref{eq:criterion-identification}) imply the convergence of $\hat{\theta
}_{n}.$ By identical arguments (under $\mathsf{H}_{0}$) we can prove that
$\tilde{\theta}_{n}$ converges, replacing $\Theta$ by $\Theta_{\mathsf{H}_{0}%
}$.\hfill\ $\square$

\subsubsection*{\emph{Proof that Assumption \ref{ass:derivatives} holds}}

Note that
\begin{align*}
&  \sup_{\theta\in\Theta:\Vert\theta-\theta_{n}\Vert\leq\epsilon_{n}%
}\left\Vert n^{-1}\frac{\partial^{2}L_{n}(\theta)}{\partial\theta
\partial\theta^{\prime}}-n^{-1}\frac{\partial^{2}L_{n}(\theta_{n})}%
{\partial\theta\partial\theta^{\prime}}\right\Vert \\
&  \leq2\sup_{\theta\in\Theta}\left\Vert n^{-1}\frac{\partial^{2}L_{n}%
(\theta)}{\partial\theta\partial\theta^{\prime}}-\mathbb{E}_{\psi_{0}}\left[
\frac{\partial^{2}l_{t}(\theta)}{\partial\theta\partial\theta^{^{\prime}}%
}\right]  \right\Vert \\
&  +\sup_{\theta\in\Theta:\Vert\theta-\theta_{n}\Vert\leq\epsilon_{n}%
}\left\Vert \mathbb{E}_{\psi_{0}}\left[  \frac{\partial^{2}l_{t}(\theta
)}{\partial\theta\partial\theta^{^{\prime}}}\right]  -\mathbb{E}_{\psi_{0}%
}\left[  \frac{\partial^{2}l_{t}(\theta_{n})}{\partial\theta\partial
\theta^{^{\prime}}}\right]  \right\Vert .
\end{align*}
Hence Assumption \ref{ass:derivatives} holds provided%
\begin{equation}
\sup_{\theta\in\Theta}\left\Vert n^{-1}\frac{\partial^{2}L_{n}(\theta
)}{\partial\theta\partial\theta^{\prime}}-\mathbb{E}_{\psi_{0}}\left[
\frac{\partial^{2}l_{t}(\theta)}{\partial\theta\partial\theta^{^{\prime}}%
}\right]  \right\Vert =o_{p}(1), \label{eq:sup_ARCH_hessian}%
\end{equation}
and that $\mathbb{E}_{\psi_{0}}[\partial^{2}l_{t}(\theta)/\partial
\theta\partial\theta^{^{\prime}}]$ is continuous. Both conditions are shown by
an application of Lemma 11.3 of Andrews and Cheng (2013b), and the proof
follows closely the arguments given in the previous proof. Note initially,
that (\ref{eq:sup_ARCH_hessian}) holds provided that for any $i,j=1,\dots
,d_{\theta}$
\begin{equation}
\sup_{\theta\in\Theta}\left\vert n^{-1}\frac{\partial^{2}L_{n}(\theta
)}{\partial\theta_{i}\partial\theta_{j}}-\mathbb{E}_{\psi_{0}}\left[
\frac{\partial^{2}l_{t}(\theta)}{\partial\theta_{i}\partial\theta_{j}}\right]
\right\vert =o_{p}(1). \label{eq:sup_ARCH_hessian_ij}%
\end{equation}
It holds that for any $i,j=1,\dots,d_{\theta}$,%
\begin{align*}
\frac{\partial^{2}l_{t}(\theta)}{\partial\theta_{i}\partial\theta_{j}}  &
=-\frac{1}{2}\left[  2\frac{y_{t}^{2}}{\sigma_{t}^{6}(\theta)}-\frac{1}%
{\sigma_{t}^{4}(\theta)}\right]  \left(  \frac{\partial\sigma_{t}^{2}(\theta
)}{\partial\theta_{i}}\right)  \left(  \frac{\partial\sigma_{t}^{2}(\theta
)}{\partial\theta_{j}}\right) \\
&  =-\frac{1}{2}\left[  2\frac{y_{t}^{2}}{(F_{t-1}^{\prime}\theta)^{3}}%
-\frac{1}{(F_{t-1}^{\prime}\theta)^{2}}\right]  F_{t-1,i}F_{t-1,j}%
=s_{ij}(W_{t},\theta),
\end{align*}
with%
\[
s_{ij}(w,\theta)=-\frac{1}{2}\left[  2\frac{w_{1}}{(w_{2}^{\prime}\theta)^{3}%
}-\frac{1}{(w_{2}^{\prime}\theta)^{2}}\right]  w_{2,i}w_{2,j},\quad
w=(w_{1},w_{2}^{\prime})^{\prime}\in\mathcal{W}\text{.}%
\]
For any $\theta_{1},\theta_{2}\in\Theta$, a mean value expansion gives that%
\[
\frac{w_{1}}{(w_{2}^{\prime}\theta_{1})^{3}}=\frac{w_{1}}{(w_{2}^{\prime
}\theta_{2})^{3}}-3\frac{w_{1}}{(w_{2}^{\prime}\theta^{\ast})^{4}}%
w_{2}^{\prime}(\theta_{1}-\theta_{2}),
\]
and%
\[
\frac{1}{(w_{2}^{\prime}\theta_{1})^{2}}=\frac{1}{(w_{2}^{\prime}\theta
_{2})^{2}}-2\frac{1}{(w_{2}^{\prime}\theta^{\ast\ast})^{3}}w_{2}^{\prime
}(\theta_{1}-\theta_{2})
\]
with $\theta^{\ast},\theta^{\ast\ast}\in\Theta$ between $\theta_{1}$ and
$\theta_{2}$. Consequently, for any $\theta_{1},\theta_{2}\in\Theta$,%
\begin{align*}
s_{ij}(w,\theta_{1})-s_{ij}(w,\theta_{2})  &  =-\frac{1}{2}\left[
2\frac{w_{1}}{(w_{2}^{\prime}\theta_{1})^{3}}-\frac{1}{(w_{2}^{\prime}%
\theta_{1})^{2}}-\left(  2\frac{w_{1}}{(w_{2}^{\prime}\theta_{2})^{3}}%
-\frac{1}{(w_{2}^{\prime}\theta_{2})^{2}}\right)  \right]  w_{2,i}w_{2,j}\\
&  =\left[  \frac{3w_{1}}{(w_{2}^{\prime}\theta^{\ast})^{4}}-\frac{1}%
{(w_{2}^{\prime}\theta^{\ast\ast})^{3}}\right]  w_{2,i}w_{2,j}w_{2}^{\prime
}(\theta_{1}-\theta_{2}),
\end{align*}
such that%
\[
|s_{ij}(w,\theta_{1})-s_{ij}(w,\theta_{2})|\leq\underset{:=M_{ij}%
(w)}{\underbrace{\left(  \frac{3w_{1}}{\delta_{L}^{4}}+\frac{1}{\delta_{L}%
^{3}}\right)  w_{2,i}w_{2,j}\Vert w_{2}\Vert}}\Vert\theta_{1}-\theta_{2}%
\Vert.
\]
We conclude that for any $\eta>0$%
\begin{equation}
|s_{ij}(w,\theta_{1})-s_{ij}(w,\theta_{2})|\leq M_{ij}(w)\eta
\label{eq:bound_M_ij}%
\end{equation}
for all $\theta_{1},\theta_{2}\in\Theta$ and $w\in\mathcal{W}$ with
$\left\Vert \theta_{1}-\theta_{2}\right\Vert <\eta$. By Assumption ARCH
\ref{ass:ARCH-HL-criterion}, it holds that%
\begin{align*}
\mathbb{E}_{\psi}[M_{ij}(W_{t})]  &  =\mathbb{E}_{\psi}\left[  \left(
\frac{3y_{t}^{2}}{\delta_{L}^{4}}+\frac{1}{\delta_{L}^{3}}\right)
F_{t-1,i}F_{t-1,j}\Vert F_{t-1}\Vert\right] \\
&  \leq\frac{3}{\delta_{L}^{4}}\mathbb{E}_{\psi}\left[  y_{t}^{2}\left\Vert
F_{t-1}\right\Vert ^{3}\right]  +\delta_{L}^{-3}\mathbb{E}_{\psi}\left[
\left\Vert F_{t-1}\right\Vert ^{3}\right]  \leq\tilde{c}%
\end{align*}
for some constant $\tilde{c}\in(0,\infty)$ for all $\psi\in\Psi$. Moreover, we
have that for $\theta\in\Theta$%
\[
|s_{ij}(W_{t},\theta)|\leq\frac{1}{2}\left[  2\frac{y_{t}^{2}}{\delta_{L}^{3}%
}+\frac{1}{\delta_{L}^{2}}\right]  F_{t-1,i}F_{t-1,j},
\]
so applying Assumption ARCH \ref{ass:ARCH-HL-criterion} again, we have that
\[
\mathbb{E}_{\psi}\sup_{\theta\in\Theta}[|s_{ij}(W_{t},\theta)|^{1+\nu}]\leq
c^{\ast}%
\]
for some constants $\nu,c^{\ast}\in(0,\infty)$ for all $\psi\in\Psi$. We
conclude that for any $i,j=1,\dots,d_{\theta}$,
\begin{equation}
\mathbb{E}_{\psi}\sup_{\theta\in\Theta}[|s_{ij}(W_{t},\theta)|^{1+\nu
}]+\mathbb{E}_{\psi}[M_{ij}(W_{t})]\leq c, \label{eq:bound_hessian_cont}%
\end{equation}
for some constant $c\in(0,\infty)$ for all $\psi\in\Psi$. Using
(\ref{eq:bound_M_ij}) and (\ref{eq:bound_hessian_cont}) together with Lemma
11.3 of Andrews and Cheng (2013b), we have that (\ref{eq:sup_ARCH_hessian_ij})
holds and, hence, that (\ref{eq:sup_ARCH_hessian}) holds. Moreover, this lemma
ensures that $\mathbb{E}_{\psi_{0}}[\partial^{2}l_{t}(\theta)/\partial
\theta_{i}\partial\theta_{j}]$ is uniformly continuous on $\Theta$ for all
$\psi_{0}\in\Psi$.\hfill\ $\square$

\subsubsection*{\emph{Proof that Assumption \ref{ass:hessian} holds}}

Recall that $\Omega_{0}=-\mathbb{E}_{\psi_{0}}[\partial^{2}l_{t}(\theta
_{0})/\partial\theta\partial\theta^{\prime}]$ and note that the matrix is
positive definite for all $\psi_{0}\in\Psi$ under Assumptions ARCH
\ref{ass:ARCH-HL-stat-mixing}-\ref{ass:ARCH-HL-criterion}, by standard
arguments. It holds that%
\begin{align*}
\left\Vert -n^{-1}\frac{\partial^{2}L_{n}(\theta_{n})}{\partial\theta
\partial\theta^{\prime}}-\Omega_{0}\right\Vert  &  =\left\Vert n^{-1}%
\frac{\partial^{2}L_{n}(\theta_{n})}{\partial\theta\partial\theta^{\prime}%
}-\mathbb{E}_{\psi_{0}}\left[  \frac{\partial^{2}l_{t}(\theta_{0})}%
{\partial\theta\partial\theta^{\prime}}\right]  \right\Vert \\
&  \leq\sup_{\theta\in\Theta}\left\Vert n^{-1}\frac{\partial^{2}L_{n}(\theta
)}{\partial\theta\partial\theta^{\prime}}-\mathbb{E}_{\psi_{0}}\left[
\frac{\partial^{2}l_{t}(\theta)}{\partial\theta\partial\theta^{\prime}%
}\right]  \right\Vert \\
&  +\left\Vert \mathbb{E}_{\psi_{0}}\left[  \frac{\partial^{2}l_{t}(\theta
_{0})}{\partial\theta\partial\theta^{\prime}}\right]  -\mathbb{E}_{\psi_{0}%
}\left[  \frac{\partial^{2}l_{t}(\theta_{n})}{\partial\theta\partial
\theta^{\prime}}\right]  \right\Vert ,
\end{align*}
where the first term is $o_{p}(1)$ for all $\psi_{0}\in\Psi$, by the arguments
given in the proof of Assumption \emph{\ref{ass:derivatives}. }Moreover, from
that proof, it holds that the second term is $o(1)$ for all $\psi_{0}\in\Psi
$.\hfill\ $\square$

\subsubsection*{\emph{Proof that Assumption \ref{ass:score} holds}}

For a given $\psi_{n}$, the (scaled) score is given by%
\[
S_{n}=\frac{1}{\sqrt{n}}\sum_{t=1}^{n}\frac{\partial l_{t}(\theta_{n}%
)}{\partial\theta}=\frac{1}{\sqrt{n}}\sum_{t=1}^{n}-\frac{1}{2}(\varepsilon
_{t}^{2}-1)\frac{1}{\sigma_{t}^{2}(\theta_{n})}F_{t-1},
\]
with $F_{t}$ being $\mathcal{F}_{t,n}$-measurable. For any non-zero constant
vector $k\in\mathbb{R}^{d_{\theta}}$, let $s_{t}=-k^{\prime}F_{t-1}\sigma
_{t}^{-2}(\theta_{n})(\varepsilon_{t}^{2}-1)/2$, such that $k^{\prime}%
S_{n}=n^{-1/2}\sum_{t=1}^{n}s_{t}$, and note that by Assumption ARCH
\ref{ass:ARCH-HL-innovation}, $\mathbb{E}_{\psi_{n}}[s_{t}|\mathcal{F}%
_{t-1,n}]=0$ almost surely. The result follows by an application of the
Lindeberg CLT for martingale difference arrays combined with an application of
the Cram\'{e}r-Wold Theorem. Note that by Assumptions ARCH
\ref{ass:ARCH-HL-innovation} and ARCH \ref{ass:ARCH-HL-criterion},
\begin{align*}
\mathbb{E}_{\psi_{n}}[s_{t}^{2}]  &  =\mathbb{E}_{\psi_{n}}[(\varepsilon
_{t}^{2}-1)^{2}/4]k^{\prime}\mathbb{E}_{\psi_{n}}[\sigma_{t}^{-4}(\theta
_{n})F_{t-1}F_{t-1}^{\prime}]k\\
&  =(\kappa/4)k^{\prime}\mathbb{E}_{\psi_{n}}[\sigma_{t}^{-4}(\theta
_{n})F_{t-1}F_{t-1}^{\prime}]k.
\end{align*}
The convergence of parameters under the drifting sequence induces convergence
in distribution of $(F_{t-1},\theta_{n})$ under $\psi_{n}$ to $(F_{t-1}%
,\theta_{0})$ under $\psi_{0}$. Consequently, by the continuous mapping
theorem $\sigma_{t}^{-4}(\theta_{n})F_{t-1}F_{t-1}^{\prime}=(\theta
_{n}^{\prime}F_{t-1})^{-2}F_{t-1}F_{t-1}^{\prime}$ converges in distribution
to $(\theta_{0}^{\prime}F_{t-1})^{-2}F_{t-1}F_{t-1}^{\prime}$ under $\psi_{0}%
$. Assumption ARCH \ref{ass:ARCH-HL-criterion} implies that $\Vert
F_{t-1}\sigma_{t}^{-2}(\theta_{n})\Vert^{2}\leq\delta_{L}^{-2}\Vert
F_{t-1}\Vert^{2}$ is uniformly integrable, and consequently, we have that
$\mathbb{E}_{\psi_{n}}[\sigma_{t}^{-4}(\theta_{n})F_{t-1}F_{t-1}^{\prime
}]\rightarrow\mathbb{E}_{\psi_{0}}[\sigma_{t}^{-4}(\theta_{0})F_{t-1}%
F_{t-1}^{\prime}]$. Hence,%
\[
\mathbb{E}_{\psi_{n}}[s_{t}^{2}]\rightarrow k^{\prime}\Sigma_{0}k,
\]
with%
\[
\Sigma_{0}:=\left(  \kappa/4\right)  \mathbb{E}_{\psi_{0}}[\sigma_{t}%
^{-4}(\theta_{0})F_{t-1}F_{t-1}^{\prime}].
\]
The matrix $\Sigma_{0}$ is positive definite by standard arguments and
Assumption ARCH \ref{ass:identification}. It remains to show that for any
constant $\epsilon>0$,%
\begin{equation}
\frac{1}{n}\sum_{t=1}^{n}s_{t}^{2}\mathbb{I}(s_{t}^{2}>\sqrt{n}\epsilon
)\overset{p}{\rightarrow}0, \label{eq:ARCH:Lindeberg}%
\end{equation}
and%
\begin{equation}
\frac{1}{n}\sum_{t=1}^{n}\mathbb{E}\left[  s_{t}^{2}|\mathcal{F}%
_{t-1,n}\right]  -\mathbb{E}_{\psi_{n}}[s_{t}^{2}]\overset{p}{\rightarrow}0.
\label{eq:ARCH:variance_convergence}%
\end{equation}
To show (\ref{eq:ARCH:Lindeberg}), note that by Assumptions ARCH
\ref{ass:ARCH-HL-innovation} and ARCH \ref{ass:ARCH-HL-criterion} and
Minkowski's inequality there exist constants a $\upsilon,c\in(0,\infty)$ (with
$c$ depending on $k$) such that for any $\eta>0,$%
\begin{align*}
&  \mathbb{P}_{\psi_{n}}\left(  \frac{1}{n}\sum_{t=1}^{n}s_{t}^{2}%
\mathbb{I}(s_{t}^{2}>\sqrt{n}\epsilon)>\eta\right)  \leq\eta^{-1}%
\mathbb{E}_{\psi_{n}}\left[  s_{t}^{2}\mathbb{I}(s_{t}^{2}>\sqrt{n}%
\epsilon)\right] \\
&  \leq\frac{1}{\eta\left(  \sqrt{n}\epsilon\right)  ^{\upsilon}}%
\mathbb{E}_{\psi_{n}}\left[  |s_{t}|^{2+\upsilon}\right]  =\frac{1}%
{\eta\left(  \sqrt{n}\epsilon\right)  ^{\upsilon}}\mathbb{E}_{\psi_{n}}\left[
|(\varepsilon_{t}^{2}-1)\sigma_{t}^{-2}(\theta_{n})k^{\prime}F_{t-1}%
/2|^{2+v}\right] \\
&  =\frac{1}{2^{2+\upsilon}\eta\left(  \sqrt{n}\epsilon\right)  ^{\upsilon}%
}\mathbb{E}_{\psi_{n}}\left[  |\varepsilon_{t}^{2}-1|^{2+\upsilon}\right]
\mathbb{E}_{\psi_{n}}\left[  |\sigma_{t}^{-2}(\theta_{n})k^{\prime}%
F_{t-1}|^{2+\upsilon}\right] \\
&  \leq\frac{1}{2^{2+\upsilon}\eta\left(  \sqrt{n}\epsilon\right)  ^{\upsilon
}}\mathbb{E}_{\psi_{n}}\left[  |\varepsilon_{t}^{2}-1|^{2+\upsilon}\right]
\delta_{L}^{-(2+v)}\left(  \sum_{i=1}^{d_{\theta}}|k_{i}|(\mathbb{E}_{\psi
_{n}}\left[  \Vert F_{t-1}\Vert^{2+\upsilon}\right]  )^{1/(2+\upsilon
)}\right)  ^{2+\upsilon}\\
&  \leq\frac{1}{2^{2+\upsilon}\eta\left(  \sqrt{n}\epsilon\right)  ^{\upsilon
}}c\rightarrow0,
\end{align*}
and we conclude that (\ref{eq:ARCH:Lindeberg}) holds. The convergence in
(\ref{eq:ARCH:variance_convergence}) follows by an application of the (weak)
LLN for row-wise stationary and strongly mixing triangular arrays (Andrews,
1988, p.~462), using that $\mathbb{E}_{\psi_{n}}\left[  s_{t}^{2}%
|\mathcal{F}_{t-1,n}\right]  =(\kappa/4)(\sigma_{t}^{-2}(\theta_{n})k^{\prime
}F_{t-1})^{2}$ is uniformly integrable under Assumption ARCH
\ref{ass:ARCH-HL-criterion}.\hfill\ $\square$

\subsubsection*{\emph{Proof that Assumption \ref{ass:covariance-consistency}
holds}}

First, note that by Assumptions \ref{ass:consistency}--\ref{ass:hessian},
$\hat{\Omega}_{n}\overset{p}{\rightarrow}\Omega_{0}$. Consequently, it remains
to show that
\begin{equation}
\hat{\kappa}_{n}\overset{p}{\rightarrow}\kappa. \label{eq:convergence_kappa}%
\end{equation}
We have that%
\[
\hat{\kappa}_{n}=n^{-1}\sum_{t=1}^{n}\left(  \frac{y_{t}^{4}}{(\hat{\theta
}_{n}^{\prime}F_{t-1})^{2}}-1\right)  .
\]
Using the same notation as before, let $f:\mathcal{W\times\Theta\rightarrow
}\mathbb{R}$ be given by%
\[
f(w,\theta)=\frac{w_{1}^{2}}{\left(  \theta^{\prime}w_{2}\right)  ^{2}},
\]
such that $\hat{\kappa}_{n}=n^{-1}\sum_{t=1}^{n}f(W_{t},\hat{\theta}_{n})$.
With $\theta_{1},\theta_{2}\in\Theta$, a mean-value expansion gives that
\[
f(w,\theta_{1})-f(w,\theta_{2})=-2\frac{w_{1}^{2}}{(\theta^{\ast\prime}%
w_{2})^{3}}w_{2}^{\prime}(\theta_{1}-\theta_{2}).
\]
It holds that%
\[
|-2\frac{w_{1}^{2}}{(\theta^{\ast\prime}w_{2})^{3}}w_{2}^{\prime}(\theta
_{1}-\theta_{2})|\leq\frac{2w_{1}^{2}}{\delta_{L}^{3}}\Vert w_{2}\Vert
\Vert\theta_{1}-\theta_{2}\Vert
\]
such that with $M:\mathcal{W\rightarrow}\mathbb{R}$ given by%
\[
M(w)=\frac{2w_{1}^{2}}{\delta_{L}^{3}}\Vert w_{2}\Vert,
\]%
\[
|f(w,\theta_{1})-f(w,\theta_{2})|\leq M(w)\Vert\theta_{1}-\theta_{2}\Vert.
\]
Consequently, for any $\eta>0$%
\begin{equation}
|f(w,\theta_{1})-f(w,\theta_{2})|\leq M(w)\eta
\label{eq:bound_M_score_variance}%
\end{equation}
for all $\theta_{1},\theta_{2}\in\Theta$ and $w\in\mathcal{W}$ with
$\left\Vert \theta_{1}-\theta_{2}\right\Vert <\eta$. By Assumption ARCH
\ref{ass:ARCH-HL-criterion},
\[
\mathbb{E}_{\psi}[M(W_{t})]=\frac{2}{\delta_{L}^{3}}\mathbb{E}_{\psi}\left[
y_{t}^{2}\Vert F_{t-1}\Vert\right]  \leq\tilde{c},
\]
for some constant $\tilde{c}\in(0,\infty)$ for all $\psi\in\Psi.$ Likewise,
noting that for any $\theta\in\Theta$, $|f(W_{t},\theta)|\leq y_{t}^{4}%
/\delta_{L}^{2}$, we have by Assumption ARCH \ref{ass:ARCH-HL-criterion},%
\[
\mathbb{E}_{\psi}\left[  \sup_{\theta\in\Theta}|f(W_{t},\theta
)|^{1+\varepsilon}\right]  \leq c^{\ast}%
\]
for some constants $\varepsilon,c^{\ast}\in(0,\infty)$ for all $\psi\in\Psi$.
Consequently,
\begin{equation}
\mathbb{E}_{\psi}\left[  \sup_{\theta\in\Theta}|f(W_{t},\theta)|^{1+\nu
}\right]  +\mathbb{E}_{\psi}[M(W_{t})]\leq\bar{c}, \label{eq:bound_score_op}%
\end{equation}
for some constant $\bar{c}\in(0,\infty)$ for all $\psi\in\Psi$. Using
(\ref{eq:bound_M_score_variance}) and (\ref{eq:bound_score_op}) together with
Lemma 11.3 of Andrews and Cheng (2013b), we have that
\[
\sup_{\theta\in\Theta}\left\vert n^{-1}\sum_{t=1}^{n}f(W_{t},\theta
)-\mathbb{E}_{\psi_{0}}[f(W_{t},\theta)]\right\vert =o_{p}(1)
\]
and that $\mathbb{E}_{\psi_{0}}[f(W_{t},\theta)]$ is uniformly continuous on
$\Theta$ for all $\psi_{0}\in\Psi$. Using that $\hat{\theta}_{n}-\theta
_{0}=o_{p}(1)$, we then have that%
\[
n^{-1}\sum_{t=1}^{n}f(W_{t},\hat{\theta}_{n})-\mathbb{E}_{\psi_{0}}%
[f(W_{t},\theta_{0})]=o_{p}(1),
\]
or, equivalently, (\ref{eq:convergence_kappa}) holds.\hfill\ $\square$

\subsubsection*{Proof that Assumption \ref{ass:AN_estimator} holds}

As $\check{\beta}_{n}$ is given in terms of the Newton-Raphson estimator,
Assumption \ref{ass:AN_estimator} holds by Lemma \ref{lem:Ketz}.\hfill
\ $\square$

\section{Additional numerical results for ARCH}

In this section we provide additional simulation results, complementing the
findings in Section \ref{sec:sim_ARCH}. The data generating process for the
simulations is given by%
\begin{align*}
y_{t}  &  =\sigma_{t}z_{t},\quad t=1,\dots,n,\\
\sigma_{t}^{2}  &  =\delta_{1}+\delta_{2}y_{t-1}^{2}+\gamma x_{t-1,1}%
+\beta_{1}x_{t-1,2}+\beta_{2}x_{t-1,3}+\beta_{3}x_{t-1,4},
\end{align*}
where $\delta_{1}>0$, $\delta_{2},\gamma,\beta_{1},\beta_{2},\beta_{3}\geq0$,
and $\{z_{t}\}_{t=1,\dots,n}$ is an i.i.d. process with $z_{t}\sim N(0,1)$. We
seek to test the hypothesis%
\begin{equation}
\mathsf{H}_{0}:\gamma=0, \label{eq null}%
\end{equation}
against $\gamma>0$.

In terms of the covariates, we let
\[
x_{i,t}=\frac{\tilde{X}_{i,t}}{\mathbb{E}[\tilde{X}_{i,t}]},\quad
i=1,\dots,4,\quad t=1,\dots,T,
\]
where
\[
\tilde{X}_{i,t}=F_{i}^{-1}(U_{i,t}),\quad i=1,\dots,4,
\]
with $F_{i}^{-1}(\cdot)$ the inverse distribution function of $\Gamma
(a_{i},b)$ for $i=1,2,3$, $b=10$ and $a_{1}=3,a_{2}=5,a_{3}=10,$ and
$F_{4}^{-1}(\cdot)$ is the inverse distribution function of $\chi_{5}^{2}$.
The correlated uniform variables $U_{i,t}=\Phi(Z_{i,t})$ for $i=1,\dots,4,$
where $(Z_{t})_{t=0}^{T}$ is an i.i.d. process with $Z_{t}=(Z_{1,t}%
,\dots,Z_{4,t})^{\prime}\sim N_{4}(0,\Sigma)$ and $\Sigma$ a positive definite
correlation matrix.

For the experiment we assume that it is known to the researcher that the true
value of the ARCH coefficient $\delta_{2}$ is not near its boundary of zero,
so that the only parameters that potentially cause a discontinuity in the null
distribution are $\beta_{1},\beta_{2},\beta_{3}$. We report the rejection
frequencies for $n=5000$ observations, parameter values $\beta_{1}=\beta
_{2}=0,$ $\gamma,\beta_{3}\in\{0,0.01,0.05,0.1,0.25\}$ and
\[
\Sigma=%
\begin{bmatrix}
1 &  &  & \\
-0.75 & 1 &  & \\
-2/3 & 0.4 & 1 & \\
-0.1 & 0.15 & 0.35 & 1
\end{bmatrix}
,
\]
and compare with the standard LR\ as well as the CLR\ test.

Table 8 contains rejection frequencies for different values of $\beta_{3}$ and
$\gamma$.

\begin{center}
[Table 8 around here]
\end{center}

\section*{References}

\begin{description}
\item \textbf{\smallskip\noindent}\textsc{Andrews, D.W.K. }(1988):
\textquotedblleft Laws of large numbers for dependent non-identically
distributed random variables\textquotedblright, \emph{Econometric Theory},
vol. 4, 458--467.

\item \textbf{\smallskip\noindent}\textsc{Andrews, D.W.K.} (1999):
\textquotedblleft Estimation when a parameter is on a
boundary\textquotedblright, \emph{Econometrica}, vol. 67, 1341--1383.

\item \textbf{\smallskip\noindent}\textsc{Andrews, D.W.K. and Cheng, X.
}(2013),\textquotedblleft Maximum likelihood estimation and uniform inference
with sporadic identification failure\textquotedblright, \emph{Journal of
Econometrics}, vol. 173, 36--56, Supplemental Appendices.

\item \textbf{\noindent}\textsc{Andrews, D.W.K.} \textsc{and Guggenberger, P}.
(2009): \textquotedblleft Hybrid and size-corrected subsampling
methods\textquotedblright, \emph{Econometrica}, vol. 77, 721--762.

\item \textbf{\smallskip\noindent}\textsc{Andrews, D.W.K. and Guggenberger,
P.} (2010): \textquotedblleft Asymptotic size and a problem with subsampling
and with the $m$ out of $n$ bootstrap\textquotedblright, \emph{Econometric
Theory}, vol. 26, 426--468.

\item \textbf{\smallskip\noindent}\textsc{McCloskey, A. }(2017):
\textquotedblleft Bonferroni-based size-correction for nonstandard testing
problems\textquotedblright, \emph{Journal of Econometrics}, vol. 200, 17--35.

\item \textbf{\smallskip\noindent}\textsc{Silvapulle, M. J. and Sen, }P. K.
(2005). \emph{Constrained Statistical Inference: Inequality, Order and Shape
Restrictions}. John Wiley \& Sons, Hoboken, NJ.

\item \textbf{\smallskip\noindent}\textsc{White}, H. (1980): \textquotedblleft
A heteroskedasticity-consistent covariance matrix estimator and a direct test
for heteroskedasticity\textquotedblright, \emph{Econometrica}, vol. 48, 817--838.
\end{description}

\bigskip

\pagebreak\appendix

\thispagestyle{empty}%
%

\begin{table}[h]
\centering\caption{Rejection Frequencies for Different Values of $\gamma
$ and $\beta_3$}
\begin{tabular}{c|ccc}
\toprule$\beta_3$ & LR  & CLR & LR-uniform \\
\midrule\multicolumn{4}{c}{Null hypothesis, $\gamma= 0$} \\
0 & 0.1094 & 0.0497 & 0.0092 \\
0.01 & 0.1159 & 0.0468 & 0.0143 \\
0.05 & 0.1196 & 0.0497 & 0.0230 \\
0.1 & 0.1150 & 0.0484 & 0.0251 \\
0.25 & 0.1116 & 0.0456 & 0.0232 \\
\midrule\bottomrule\toprule$\gamma$ & LR & CLR & LR-uniform \\
\midrule\multicolumn{4}{c}{Alternative hypotheses with $\beta_3 = 0$} \\
0 & 0.1083 & 0.0498 & 0.0119 \\
0.01 & 0.6767 & 0.3112 & 0.2850 \\
0.05 & 1.0000 & 0.9907 & 1.0000 \\
0.1 & 1.0000 & 1.0000 & 1.0000 \\
0.25 & 1.0000 & 1.0000 & 1.0000 \\
\midrule\bottomrule\end{tabular}
\end{table}%

\end{document}